\documentclass[12pt]{article}
\usepackage{amssymb}
\usepackage{amsmath}
\usepackage{amsopn}
\usepackage{pifont}
\usepackage{color}
\usepackage{mathrsfs}
\usepackage{amsthm}

\numberwithin{equation}{section}

\theoremstyle{plain}
\newtheorem{thm}{Theorem}[section]
\newtheorem{prop}[thm]{Proposition}
\newtheorem{lemma}[thm]{Lemma}
\newtheorem{cor}[thm]{Corollary}
\newtheorem{Def}{Definition}[section]

\theoremstyle{definition}
\newtheorem{remark}{Remark}[section]

\newcommand{\mage}{\color{magenta}}

\usepackage[normalem]{ulem} 
\usepackage{cancel}

\def\bn{\begin{equation}}
\def\ed{\end{equation}}

\def\<{\langle}
\def\>{\rangle}
\newcommand{\CC}{{\mathbb C}}

\newcommand{\ZZ}{{\mathbb Z}}

\def\r#1{(\ref{#1})}

\def\ot{\otimes}

\def\sk#1{\left(#1\right)}

\def\Pep{{P}^+_e}
\def\Pem{{P}^-_e}
\def\Pepm{{P}^\pm_e}

\def\Pfpm{{P}_f^\pm}
\def\Pfp{{P}^+_f}
\def\Pfm{{P}^-_f}

\def\Uqsl2{U_q(\widehat{\mathfrak{sl}}_2)}

\def\EE{{\rm E}}\def\FF{{\rm F}}

\def\RR{{\rm R}}
\def\LL{{\rm L}}
\def\hLL{\hat{\rm L}}

\def\E{{\sf e}}

\def\tFF{\tilde{\rm F}}

\def\tEE{\tilde{\rm E}}

\def\qed{\hfill$\square$\medskip}

\def\eF{\mathcal{F}}
\def\eE{\mathcal{E}}

\def\qed{$\scriptstyle{\blacksquare}$}

\def\gle{g}
\def\gri{\tilde g}
\def\gole{\mathsf{g}}
\def\gori{\tilde{\mathsf{g}}}

\def\qed{\hfill\nobreak\hbox{$\square$}\par\medbreak}
\def\Id{\mathbb{I}}
\def\Per{\mathsf{P}}
\def\Qer{\mathsf{Q}}
\def\Der{\mathsf{D}}

\def\bFF{\bar{\rm F}}
\def\bEE{\bar{\rm E}}
\def\bk{\bar{k}}

\def\fgo{\mathsf{f}}

\def\fgs{\mathsf{f}}
\def\ggs{\mathsf{g}}
\def\gaf{\tilde{\mathfrak{g}}}
\def\gaff{\widehat{\mathfrak{g}}}
\def\AR{\mathbb{R}}
\def\QR{\mathbb{Q}}
\def\PR{\mathbb{P}}
\def\MM{\mathrm{M}}
\def\Ll{\mathbb{L}}
\def\bMM{\bar{\mathrm{M}}}
\def\qf{{\sf q}}
\def\pf{{\sf p}}

\begin{document}

\thispagestyle{empty}
\setcounter{page}{0}

\vspace{12pt}

\begin{center}
\begin{Large}
{\bf On the R-matrix realization\\[2mm] of quantum loop algebras}
\end{Large}

\vspace{20pt}

\begin{large}
A.~Liashyk${}^{a}$ and  S.~Z.~Pakuliak${}^{b}$
\end{large}

\vspace{10mm}

${}^a$ {\it Skolkovo Institute of Science and Technology, Moscow, Russia,}\\
{\it NRU Higher School of Economics, Moscow, Russia,}\\
{\it E-mail: a.liashyk@gmail.com}

\vspace{2mm}

${}^b$ {\it Bogoliubov Laboratory for Theoretical Physics, JINR, Dubna}\\
{\it Landau School of Physics and Research, NRU MIPT,}\\
{\it Dolgoprudny, Moscow region, Russia}\\
{\it E-mail: stanislav.pakuliak@jinr.int}

\end{center}


\vspace{10mm}

\noindent{\bf Abstract.}
{\footnotesize We consider $\RR$-matrix realization of the quantum deformations 
of the loop algebras $\gaf$ corresponding to non-exceptional affine Lie algebras of  type 
$\gaff=A^{(1)}_{N-1}$,  $B^{(1)}_n$, $C^{(1)}_n$, $D^{(1)}_n$, $A^{(2)}_{N-1}$.
For each $U_q(\gaf)$ we investigate the commutation relations between Gauss coordinates 
of the fundamental $\LL$-operators using embedding of the smaller 
algebra into bigger one. The new realization of these algebras 
in terms of the currents is given. 
The relations between all off-diagonal Gauss coordinates and 
certain projections from the ordered products of the currents are presented. These 
relations are important in applications to the quantum integrable models.  }


\section{Introduction}

Classification of the solutions to the quantum Yang-Baxter equation for the 
case of non-exceptional quantum affine Lie algebras was found in the pioneering paper \cite{Jimbo86}. 

Let $\mathfrak{g}$ be one of the  Lie algebras $\mathfrak{sl}_N$, $\mathfrak{o}_{2n+1}$,
$\mathfrak{sp}_{2n}$ or $\mathfrak{o}_{2n}$ corresponding to the series of the 
classical Lie algebras $A_{N-1}$, $B_n$, $C_n$ and $D_n$ respectively. 
Let $\gaff$  be one of non-exceptional affine Lie algebras $A^{(1)}_{N-1}$, 
$B^{(1)}_n$, $C^{(1)}_n$, $D^{(1)}_n$ and $A^{(2)}_{N-1}$.
By  $\gaf$ we denote the loop algebra which is  
the affine algebra $\gaff$ with zero central charge.
To save notations we will use the same names for the different 
loop algebras  $\gaf$ as for the affine algebras $\gaff$.

Let $q\in \CC$ be arbitrary complex number not equal to zero or root of unity.
In this paper we consider quantum deformation  $U_q(\gaf)$ \cite{Dr88} of the universal enveloping 
algebra  $U(\gaf)$. One may think about  $U_q(\gaf)$ 
as the corresponding  quantum affine algebra $U_q(\gaff)$ with zero central charge. 

Algebra $U_q(\gaf)$ has several descriptions. It can be formulated in terms 
of the finite number of Chevalley generators or countable set of Cartan-Weyl 
generators. Latter generators can be gathered into finite number of the 
generating series and the commutation relations between whole set of  
the Cartan-Weyl generators can be realized as finite number of the formal 
series relations between these generating series. 

For the applications to the quantum integrable models, the second description 
of $U_q(\gaf)$ is more suitable since generating series of the Cartan-Weyl 
generators can be identified with Gauss coordinates of the fundamental 
$\LL$-operators, which satisfy the same $\RR\LL\LL$-type commutation relations 
as quantum monodromies of the integrable systems do. It opens a possibility to 
construct off-shell Bethe vectors for these integrable models in terms of 
Cartan-Weyl generators of the algebra $U_q(\gaf)$ \cite{EKhP07}.

Realization of the algebra $U_q(\gaf)$ in terms of Cartan-Weyl generators 
has in turn two faces. One is given by the quadratic $\RR\LL\LL$-type commutation relations
for the fundamental $\LL$-operators defined by the  solution of the quantum
Yang-Baxter equation  \cite{Jimbo86}. This construction was first 
proposed in the paper \cite{RS1990}. On the other hand the algebra $U_q(\gaf)$
can be realized in terms of so called {\it currents} 
 \cite{D88}. For the case 
$U_q(\widehat{\mathfrak{gl}}_{N})$ an isomorphism between these two descriptions  was found in \cite{DF93}. 
Recent papers \cite{JLM19,JLM19a} prove similar 
isomorphisms  for the algebras $U_q(B^{(1)}_n)$, $U_q(C^{(1)}_n)$ and $U_q(D^{(1)}_n)$. 
Also the case $U_q(A^{(2)}_{2})$ was considered in \cite{Sh10}. 
In our investigation we extend these results to the case of $U_q(A^{(2)}_{N-1})$. 
Key observation  is the fact that $\RR$-matrix associated 
with the algebras $U_q(B^{(1)}_n)$, $U_q(C^{(1)}_n)$, $U_q(D^{(1)}_n)$, $U_q(A^{(2)}_{N-1})$
has the same structure  for all these algebras.
The differences are accumulated in one parameter $\xi$ (see~\r{Table}).

In \cite{Jimbo86} one more solution to the quantum Yang-Baxter equation was 
found. It corresponds to the affine algebra $D^{(2)}_n$. This solution has more 
complicated structure than $\RR$-matrices for above mentioned algebras.
We will describe the corresponding quantum loop algebra $U_q(D^{(2)}_n)$
in our future publications. 

The paper is composed as follows. In section~\ref{RRsect} quantum $\RR$-matrix 
for the algebra $U_q(\gaf)$ is defined together with its properties. 
Section~\ref{Aldef} is devoted to definition of the algebra 
$U_q(\gaf)$ and description of its central elements and automorphism. Gauss coordinates 
of the fundamental $\LL$-operators are introduced in section~\ref{GC}. Here we 
 discuss normal ordering 
of subalgebras in $U_q(\gaf)$ induced by the cyclic ordering of the Cartan-Weyl 
generators in the quantum affine algebras. 
Section~\ref{embed} contains the  theorem  which describes 
embedding of the smaller rank algebra $U_q(\gaf)$ into the bigger one. 
This embedding is described on the level of matrix entries of the fundamental 
$\LL$-operators and in terms of the Gauss coordinates. 
 Section~\ref{nrg} describes new realization of the algebra $U_q(\gaf)$
 in terms of the currents.  In section~\ref{curproj}
 so called composed currents are introduced
 which belong to certain completion of $U_q(\gaf)$
 and related to  
 off-diagonal Gauss coordinates of the fundamental $\LL$-operators.
 It was shown in  \cite{KhP-Kyoto,E2000} that analytical properties of the composed 
 currents and the commutation relations between them are equivalent to the 
 Serre relations between simple root currents. 
  Proofs of auxiliary propositions and lemmas are gathered in four appendices.

\section{$\RR$-matrix for $U_q(\gaf)$}
\label{RRsect}

Let $N$ be dimension of the fundamental vector representation of the algebra 
$\gaf$ in $\CC^N$. 
Let $\E_{ij}$ be a $N\times N$ matrix unit $(\E_{ij})_{k,l}=\delta_{ik}\delta_{jl}$ 
for $1\leq i,j,k,l\leq N$ and 
\begin{equation*}\label{prime}
i'=N+1-i,\quad 1\leq i\leq N\,.
\end{equation*}

To describe quantum $\RR$-matrix associated with the algebra $U_q(\gaf)$ 
\cite{Jimbo86,JLM19,JLM19a}
we define  parameter $\xi$ and {dimension}
$N$ of the fundamental vector 
representation of the algebra 
$\gaf$ given in the  table 
\begin{equation}\label{Table}
\begin{array}{|c|c|c|c|c|c|c|}
\hline
\gaf\phantom{\Big|}&A^{(1)}_{N-1}&B^{(1)}_n &C^{(1)}_n&D^{(1)}_n&A^{(2)}_{2n}&A^{(2)}_{2n-1}\\
\hline
N\phantom{\Big|}&N&2n+1&2n&2n&2n+1&2n\\
\hline
\xi\phantom{\Big|}&q^{-N}&q^{1-2n}&q^{-2-2n}&q^{2-2n}&-q^{-1-2n}&-q^{-2n}\\
\hline
\end{array}
\end{equation}
Define also the sign function  
\begin{equation*}\label{sign}
{\rm sign}(\ell)=\begin{cases}+1,\quad&\ell\geq 0\\
-1,\quad&\ell<0
\end{cases}
\end{equation*}
and a set of integers $\varepsilon_i$, $i=1,\ldots,N$
\begin{equation*}\label{signs}
\varepsilon_i=\begin{cases}
{\rm sign}(n-i),\quad &\mbox{for}\quad \gaf=C^{(1)}_{n}\,,\\
1,\quad &\mbox{for}\quad \mbox{all other cases.}
\end{cases}
\end{equation*}
For $\gaf=A^{(1)}_{N-1}, B^{(1)}_n,C^{(1)}_n,D^{(1)}_n$ and $A^{(2)}_{N-1}$ 
we need  
the map $\bar\imath$ for $i=1,\ldots,N$
\begin{equation}\label{map}
\bar\imath=\begin{cases}
\sk{\frac{N}{2}, \frac{N}{2}-1, \ldots,-\frac{N}{2}+1,-\frac{N}{2}},
\quad &\mbox{for}\quad \gaf=A^{(1)}_{N-1}\,,\\
\sk{n-\frac{1}{2},\ldots,\frac{3}{2},\frac{1}{2},0,-\frac{1}{2},-\frac{3}{2},\ldots,-n+\frac{1}{2}},
\quad &\mbox{for}\quad \gaf=B^{(1)}_n, A^{(2)}_{2n}\,,\\
(n,n-1,\ldots,1,-1,\ldots,-n),\quad &\mbox{for}\quad \gaf=C^{(1)}_{n}\,,\\
(n-1,\ldots,1,0,0,-1,\ldots,-n+1),\quad &\mbox{for}\quad \gaf=D^{(1)}_{n},A^{(2)}_{2n-1}\,.
\end{cases}
\end{equation}
Note that for any $\gaf$ we have 
\begin{equation}\label{cancel}
\bar\imath+\bar{\imath'}=0,\quad i=1,\ldots,N.
\end{equation}

 We introduce functions 
\begin{equation*}\label{rat-fun}
f(u,v)=\frac{qu -q^{-1}v}{u-v}\,,\quad 
\gle(u,v)=\frac{(q-q^{-1})u}{u-v}\,,\quad
\gri(u,v)=\frac{(q-q^{-1})v}{u-v}
\end{equation*}
of the arbitrary complex numbers  $u$ and $v$, which we call the spectral parameters.

Define matrices $\PR(u,v)$ and $\QR(u,v)$ acting in the tensor product 
$\CC^N\ot\CC^N$
\begin{equation}\label{PPuv}
\PR(u,v)= \sum_{1\leq i,j\leq N} \pf_{ij}(u,v)\ \E_{ij}\ot \E_{ji}\,,
\end{equation}
\begin{equation}\label{QQuv}
\QR(u,v)=
\sum_{1\leq i,j\leq N} \qf_{ij}(u,v)\  \E_{i'j'}\ot \E_{ij}\,,
\end{equation}
where 
rational functions $\pf_{ij}(u,v)$  and $\qf_{ij}(u,v)$ are defined as follows
\begin{equation*}\label{p-fun}
\pf_{ij}(u,v)=\begin{cases}
f(u,v)-1,\quad &i=j\,,\\
\gle(u,v),\quad &i<j\,,\\
\gri(u,v),\quad &i>j\,,
\end{cases}
\end{equation*} 
\begin{equation}\label{a-fun}
\qf_{ij}(u,v)=\varepsilon_i\varepsilon_j\ q^{\bar\imath-\bar\jmath} \begin{cases}
f(v\xi,u)-1,\quad &i=j,\quad i\not=i'\,,\\
f(v\xi,u)-1-\alpha_q,\quad &i=j,\quad i=i'\,,\\
\gle(v\xi,u),\quad &i<j\,,\\
\gri(v\xi,u),\quad &i>j\,,
\end{cases}
\end{equation}
and 
\begin{equation*}
\alpha_q=(q^{1/2}-q^{-1/2})^2\,.
\end{equation*}
 One can check that functions \r{a-fun} have a property
\begin{equation}
 q_{ij}(u,v) = q_{j' i'} (u,v).
\end{equation}

Let
\begin{equation*}\label{Id}
\Id=\sum_{i=1}^N\E_{ii}
\end{equation*}
be identity matrix in $\CC^N$.
\begin{Def}
Quantum trigonometric $\RR$-matrix acting in the tensor product of 
two fundamental vector representations of $\gaf$ 
\mbox{\rm \cite{Jimbo86}}
for the algebra $\gaf=A^{(1)}_{N-1}$ is 
\begin{equation}\label{R-matA}
\AR(u,v)=\Id\ot\Id+\PR(u,v)
\end{equation}
and for the algebras $\gaf=B^{(1)}_n$, $C^{(1)}_n$, $D^{(1)}_n$ and $A^{(2)}_{N-1}$ is 
\begin{equation}\label{R-mat}
\RR(u,v)=\AR(u,v)+\QR(u,v)=\Id\otimes\Id + \PR(u,v)+\QR(u,v).
\end{equation}
\end{Def}

For any $X\in{\rm End}(\CC^N)$  transposed matrix $X^{\rm t}$ 
is
\begin{equation}\label{trans}
(X^{\rm t})_{i,j}=\varepsilon_i\, \varepsilon_j\, X_{j',i'}\,.
\end{equation} 
Let $\Der$ be a diagonal matrix 
\begin{equation*}\label{Der}
\Der={\rm diag}(q^{\bar 1},q^{\bar 2},\ldots,q^{\bar N}),
\end{equation*}
where $\bar\imath$ for $i=1,\ldots,N$ are given by \r{map}.

Let $\Per$  be permutation operator
($\Per^2=\Id$) in $\CC^N\ot\CC^N$ and  $\Qer$ be projector ($\Qer^2=N\Qer$)
onto one-dimensional subspace in $\CC^N\ot\CC^N$
\begin{equation*}\label{PQer}
\Per=\sum_{1\leq i,j\leq N\atop} \E_{ij}\ot \E_{ji},\quad 
\Qer=\sum_{1\leq i,j\leq N\atop} \varepsilon_i\varepsilon_j\E_{i'j'}\ot \E_{ij}=\Per^{{\rm t}_1}
=\Per^{{\rm t}_2}\,.
\end{equation*}

Trigonometric $\gaf$-invariant $\RR$-matrix given by \r{R-matA} and \r{R-mat} 
possesses following properties.
\begin{itemize}
\item {\it Scaling invariance}
\begin{equation}\label{scaling}
\RR(\beta u,\beta v)=\RR(u,v)
\end{equation}
 for any complex parameter $\beta$ which is not equal to zero.
\item {\it Transposition symmetry} 
 \begin{equation}\label{reflec}
 \RR_{12}(u,v)^{{\rm t}_1{\rm t}_2}=\RR_{12}(u,v)\,.
 \end{equation}
\item {\it Twist symmetry}
 \begin{equation}\label{comm}
 K_1\ K_2\ \RR_{12}(u,v)=\RR_{12}(u,v)\ K_1\ K_2\,,
 \end{equation}
 where $K$ is $N\times N$ $\CC$-valued matrix such that $K K^{\rm t} = \Id$, 
 $K_1=K\ot\Id$ and $K_2=\Id\ot K$. Due to \r{cancel} $\Der^{{\rm t}}=\Der^{-1}$ 
 and equality \r{comm} 
 is valid for $K=\Der$. 
 \item {\it Yang-Baxter equation}
 \begin{equation}\label{YB}
\RR_{12}(u,v)\cdot \RR_{13}(u,w)\cdot \RR_{23}(v,w)=
\RR_{23}(v,w)\cdot \RR_{13}(u,w)\cdot \RR_{12}(u,v)\,,
\end{equation}
where subscripts of $\RR$-matrices mean the indices of the spaces $\CC^N$ where 
it acts nontrivially.
 \item {\it Unitarity}
\begin{equation}\label{unitar}
\RR_{12}(u,v)\cdot \RR_{21}(v,u)=f(u,v)f(v,u)\ \Id\ot\Id\,,
\end{equation}
where $\RR_{21}(u,v)=\Per_{12}\,\RR_{12}(u,v)\,\Per_{12}$.
\item {\it Crossing type symmetries} 
\begin{equation}\label{crossingA}
\Der^2_1\ \AR_{12}(v\xi^2,u)^{{\rm t}_1}\ \Der_1^{-2}\ \AR_{21}(u,v)^{{\rm t}_1}=\Id\ot\Id
\end{equation} 
for $\gaf=A^{(1)}_{N-1}$  with $\xi=q^{-N}$  and 
\begin{equation}\label{crossing}
 \Der_1\ \RR_{12}(v\xi,u)^{{\rm t}_1}\ \Der_1^{-1}\ \RR_{12}(v,u)
=f(u,v)f(v,u)\ \Id\ot\Id
\end{equation} 
for $\gaf=B^{(1)}_n,C^{(1)}_n,D^{(1)}_n,A^{(2)}_{N-1}$.
Crossing relation \r{crossing} follows from the presentation of the matrix $\QR(u,v)$
given by \r{QQuv} in the form 
\begin{equation*}\label{Qerp}
\QR(u,v)= \Der_2\ \Per_{12}\ \PR_{12}(v\xi,u)^{{\rm t}_1}\ \Per_{12}\ \Der_2^{-1}-
\alpha_q\delta_{N,{\rm odd}}\ \E_{n+1,n+1}\ot\E_{n+1,n+1}
\end{equation*}
where $\delta_{N,{\rm odd}}=1$ for $N=2n+1$ and 0 for $N=2n$. 
This relation implies 
\begin{equation*}\label{Rrel1}
\RR_{12}(u,v)=\Der_2\ \Per_{12}\ \RR_{12}(v\xi,u)^{{\rm t}_1}\ \Per_{12}\ \Der_2^{-1}
\end{equation*} 
which is equivalent to \r{crossing} due to \r{unitar}. Crossing symmetry \r{crossing} 
for $\RR$-matrix  \r{R-mat} yields the relation similar to \r{crossingA}
\begin{equation}\label{Cross}
 \Der^2_1\ \RR_{12}(v\xi^2,u)^{{\rm t}_1}\ \Der_1^{-2}\ \RR_{21}(u,v)^{{\rm t}_1}
=f(u,v\xi)f(v\xi,u)\ \Id\ot\Id
\end{equation} 
\item {\it Pole structure}\\
$\RR$-matrix \r{R-matA} for $\gaf=A^{(1)}_{N-1}$ has simple pole at $u=v$ 
 \begin{equation}\label{poleA}
\left.\frac{(u-v)}{u(q-q^{-1})}\ \AR_{12}(u,v)\right|_{u=v}=\Per_{12},\quad
\end{equation} 
while
$\RR$-matrix \r{R-mat} for 
$\gaf=B^{(1)}_n,C^{(1)}_n,D^{(1)}_n,A^{(2)}_{N-1}$
has two simple poles at $u=v$ and $u=v\xi$ with residues
\begin{equation}\label{poles}
\left.\frac{(u-v)}{u(q-q^{-1})}\ \RR_{12}(u,v)\right|_{u=v}=\Per_{12},\quad
\left.\frac{(v\xi-u)}{u(q-q^{-1})}\ \RR_{12}(u,v)\right|_{u=v\xi}=\Der_1^{-1}\,\Per_{12}^{{\rm t}_1}
\,\Der_1.
\end{equation}
 This pole structure and crossing relations \r{crossingA}  and \r{Cross} allow 
to get for both $\RR$-matrices \r{R-matA} and \r{R-mat}
\begin{equation}\label{cen2nonA}
x(u, v\xi^2)^{-1}
\left.\frac{v\xi^2 -u}{u(q-q^{-1})}\Big(\RR_{12}(u,v)^{{\rm t}_1}\Big)^{-1}\right|_{u=v\xi^2}=
 \Der_1^{-2}\ \Per_{12}^{{\rm t}_1}\ \Der_1^{2}\,,
\end{equation}
 where rational function $x(u,v)$ is 
 \begin{equation}\label{xuv}
x(u,v)=\begin{cases}
1,\qquad&\mbox{for}\quad \gaf=A^{(1)}_{N-1}\,,\\
f(u\xi,v)^{-1}f(v,u\xi)^{-1}&\mbox{for all other}\quad \gaf\,.
\end{cases}
\end{equation}
 
\item {\it Scaling limit}\\
In the scaling limit $\epsilon\to 0$ and 
$u\to e^{\epsilon u}$, $v\to e^{\epsilon v}$, $q\to e^{\epsilon c/2}$, $\xi\to
e^{-\epsilon c\kappa}$
trigonometric $R$-matrix \r{R-mat}   goes into rational $\mathfrak{g}$-invariant 
$R$-matrix
\begin{equation}\label{R-rat}
\RR(u,v)\ =\  \Id\ot \Id+\frac{c}{u-v}\ \Per\ -
\frac{c}{u-v+c\kappa}\ \Qer
\end{equation}
for the algebras $\gaf=B^{(1)}_n,C^{(1)}_n,D^{(1)}_n$ 
and into rational $\mathfrak{gl}_N$-invariant  the $\RR$-matrix 
\begin{equation*}\label{R-rat-A}
\AR(u,v)\ =\  \Id\ot \Id+\frac{c}{u-v}\ \Per
\end{equation*}
for the algebras $\gaf=A^{(1)}_{N-1},A^{(2)}_{N-1}$.
\end{itemize}
Quantum $\RR$-matrix \r{R-rat} appeared in investigation of the classical series 
 Yangians and their doubles in  \cite{JLM18,JLY18,LP20}.

\section{$\RR$-matrix formulation of the algebra $U_q(\gaf)$}
\label{Aldef}

The algebra $U_q(\gaf)$ over $\CC(q)$ 
(over $\CC(q^{1/2})$ for $\gaf=B^{(1)}_n$ and $A^{(2)}_{2n}$)
is generated by the elements $\LL^\pm_{i,j}[\pm m]$,
$1\leq i,j\leq N$, $m\in\ZZ_+$ such that 
\begin{equation}\label{restr}
\LL^+_{j,i}[0]=\LL^-_{i,j}[0]=0, \quad i<j,\quad \LL^+_{i,i}[0]\LL^-_{i,i}[0]=\LL^-_{i,i}[0]\LL^+_{i,i}[0]=1\,.
\end{equation}
There are also 
additional relations for the  operators $\LL^\pm_{i,j}[\pm m]$ which 
are due to existence of the central elements in $U_q(\gaf)$ described in section~\ref{cenel31}.

The generators of the algebra $U_q(\gaf)$ can be gathered into formal series 
\begin{equation}\label{series}
\LL^\pm_{i,j}(u)=\sum_{m=0}^\infty \LL^\pm_{i,j}[\pm m] u^{\mp m}
\end{equation}
and combined in the matrices 
\begin{equation}\label{Lop}
\LL^\pm(u)=\sum_{i,j=1}^N \E_{ij}\ot \LL^\pm_{i,j}(u)\in {\rm End}(\CC^N)\ot 
U_q(\gaf)[[u,u^{-1}]]
\end{equation}
which we call $\LL$-operators\footnote{Further on we will skip the sign $\ot$ of the tensor 
product in \r{Lop} and write simple $\LL^\pm(u)=\sum_{i,j}\E_{ij}\LL^\pm_{i,j}(u)$.}. 
The commutation relations in the algebra $U_q(\gaf)$ are given 
by the standard RLL commutation relations in $(\CC^N)^{\ot2}\ot U_q(\gaf)[[u,u^{-1}]]$
\begin{equation}\label{RLL}
\RR(u,v) \cdot (\LL^\mu(u) \ot \Id)\cdot (\Id\ot \LL^\nu(v))=
(\Id\ot \LL^\nu(v))\cdot (\LL^\mu(u) \ot \Id)\cdot \RR(u,v)\,,
\end{equation}
where $\mu,\nu=\pm$  and rational functions entering $\RR$-matrices \r{R-matA} and \r{R-mat} 
 should be understood as series over $v/u$ for $\mu=+$, $\nu=-$ and 
as series over $u/v$ for $\mu=-$, $\nu=+$. For $\mu=\nu$  
these rational functions can be either series over the ratio $v/u$ or the ratio  $u/v$.

The commutation relations in the algebra $U_q(\gaf)$ may be written in terms
of matrix entries \r{series}. Using explicit expression 
\r{PPuv} and \r{QQuv} one  gets
\begin{equation}\label{TM-1}
\begin{split}
&[\LL^\mu_{i,j}(u),\LL^\nu_{k,l}(v)]=
\pf_{lj}(u,v)\ \LL^\nu_{k,j}(v)\LL^\mu_{i,l}(u)-\pf_{ik}(u,v)\ 
\LL^{\mu}_{k,j}(u)\LL^{\nu}_{i,l}(v)+ \\
&\qquad+\sum_{p=1}^N \sk{\delta_{l,j'}\  \qf_{pl}(u,v)\ 
\LL^\nu_{k,p}(v)\LL^\mu_{i,p'}(u)-
\delta_{i,k'}\  \qf_{kp}(u,v)\ 
\LL^\mu_{p',j}(u)\LL^\nu_{p,l}(v)}\,.
\end{split}
\end{equation}
The sum in the last line 
of the commutation relations \r{TM-1} is absent for the algebra $\gaf=A^{(1)}_{N-1}$. 
It follows from the commutation relations \r{RLL} or \r{TM-1} that modes $\LL^+_{i,j}[m]$ 
and $\LL^-_{i,j}[-m]$, $m\geq 0$ form Borel subalgebras $U^\pm_{q}(\gaf)\subset U_q(\gaf)$.

\begin{remark}\label{rem1}
One can  check that the restrictions to the 
zero mode generators \r{restr} are consistent  with the commutation relations 
\r{TM-1}. 
Indeed, taking the limit $u\to\infty$ in \eqref{TM-1} with $\mu = +$ and 
using the expansion \eqref{series}, one gets
\begin{equation}\label{TM-zm}
\begin{split}
&q^{\delta_{ik}}\LL^+_{i,j}[0]\,\LL^\nu_{k,l}(v)-q^{\delta_{jl}}\LL^\nu_{k,l}(v)\,\LL^+_{i,j}[0]=\\
&\qquad=(q-q^{-1})
\Big(\delta_{l<j}\ \LL^\nu_{k,j}(v)\LL^+_{i,l}[0]-\delta_{i<k}\ \LL^+_{k,j}[0]\LL^\nu_{i,l}(v)\Big)+\\
&\quad\qquad +
\sum_{p=1}^N \Big(\delta_{l,j'} \qf_{pl}
\LL^\nu_{k,p}(v)\LL^+_{i,p'}[0]-
\delta_{i,k'} \qf_{kp}
\LL^+_{p',j}[0]\LL^\nu_{p,l}(v)\Big),
\end{split}
\end{equation}
where $\delta_{i<j}=1$ if $i<j$ and 0 otherwise and 
\begin{equation*}\label{a-fun-con}
\qf_{ij}=\varepsilon_i\varepsilon_j\ q^{\bar\imath-\bar\jmath}\begin{cases}
q^{-1}-1,\quad &i=j,\quad i\not=i' {\mage ,} \\
1-q,\quad &i=j,\quad i=i'\,,\\
0,\quad &i<j\,,\\
-q^{\bar\imath-\bar\jmath}(q-q^{-1}),\quad &i>j\,.
\end{cases}
\end{equation*}

Now, if one supposes that $i>j$ and applying \r{restr} for 
the zero mode operators $\LL^+_{i,j}[0]$, the l.h.s. of \r{TM-zm} vanishes identically. 
Due to the coefficients $\delta_{l<j}$ and $\delta_{i<k}$ in the second line of 
\r{TM-zm} and the combinations $\qf_{pj'}\LL^+_{i,p'}[0]$, $\qf_{i'p}\LL^+_{p',j}[0]$
in the third line of this equality,  the r.h.s. also vanishes for the same reason.

Analogously, one can check that the restriction that zero mode operators $\LL^-_{i,j}[0]$ vanishes 
for $i<j$ is consistent with the  series expansion \r{series} in $u$ of the $L$-operator 
$\LL^-(u)$. 
In that case, the zero modes occur in the limit $u\to0$, which changes the exchange 
relations \r{TM-zm} 
and makes everything consistent again. 
Finally, one can also verify that the limit $v\to\infty$ in \r{TM-1} for 
$\LL^+_{k,l}(v)$ and the limit $v\to0$ for $\LL^-_{k,l}(v)$ 
leads to the same conclusions.
\end{remark}

\subsection{Central elements in $U_q(\gaf)$}
\label{cenel31}

Due to the commutation relations \r{RLL} algebra $U_q(\gaf)$ has central elements
which are given by the following 
\begin{prop}
For any $\gaf=A^{(1)}_{N-1}$, $A^{(2)}_{N-1}$, $B^{(1)}_n$, $C^{(1)}_n$ and $D^{(1)}_n$ 
the algebra
$U_q(\gaf)$ has the central elements $Z^\pm(v)\in U^\pm_q(\gaf)$
\begin{equation}\label{Cen1}
Z^\pm(v)\ \Id=
\Der^{2}\  \LL^\pm(v\xi^2)^{{\rm t}}\ \Der^{-2}\ \Big(\LL^\pm(v)^{-1}\Big)^{{\rm t}}=
\Big(\LL^\pm(v)^{-1}\Big)^{{\rm t}}\ \Der^{2}\ \LL^\pm(v\xi^2)^{{\rm t}}\ \Der^{-2},
\end{equation}
where parameter $\xi$ is given by the table~\r{Table}.
 Equation \eqref{Cen1} means that 
products of the matrices 
\begin{equation*}
 \Der^{2}\  \LL^\pm(v\xi^2)^{{\rm t}}\ \Der^{-2}\ \Big(\LL^\pm(v)^{-1}\Big)^{{\rm t}} \quad
\text{ and }
\quad \Big(\LL^\pm(v)^{-1}\Big)^{{\rm t}}\ \Der^{2}\ \LL^\pm(v\xi^2)^{{\rm t}}\ \Der^{-2}
\end{equation*}
are  equal and proportional to the unit matrix $\Id$.  
The proportionality coefficients are the central elements.
\end{prop}

{\it Proof.} To find central elements \r{Cen1} one can
transform the commutation relations for the 
fundamental $\LL$-operators  \r{RLL} to the form\footnote{In what follows  
we will sometimes skip superscripts 
of $\LL$-operators. If these superscripts is not explicitly mentioned it means that the corresponding 
relation is valid for both values $\pm$.} 
\begin{equation*}\label{cen1}
\big(\RR_{12}(u,v)^{{\rm t}_1}\big)^{-1}\ \LL^{(2)}(v)^{-1}\ \LL^{(1)}(u)^{{\rm t}_1}=
\LL^{(1)}(u)^{{\rm t}_1}\ \LL^{(2)}(v)^{-1}\ \big(\RR_{12}(u,v)^{{\rm t}_1}\big)^{-1},
\end{equation*}
where standard notations 
\begin{equation*}\label{st-not}
\LL^{(1)}(u)=\LL(u)\ot\Id,\qquad \LL^{(2)}(u)=\Id\ot\LL(u)
\end{equation*}
are used.
Taking the residue at the point $u=v\xi^2$ in this equation 
and using \r{cen2nonA} one gets
\begin{equation*}\label{cen3}
\Der_1^{-2}\ \Per_{12}^{{\rm t}_1}\ \Der_1^{2}\ \LL^{(2)}(v)^{-1}\ \LL^{(1)}(v\xi^2)^{{\rm t}_1}=
\LL^{(1)}(v\xi^2)^{{\rm t}_1}\ \LL^{(2)}(v)^{-1}\ \Der_1^{-2}\ \Per_{12}^{{\rm t}_1}\ \Der_1^{2}
\end{equation*}
or 
\begin{equation*}\label{cen5}
\Big(\LL^{(1)}(v)^{-1}\Big)^{{\rm t}_1}\ \Der_1^{2}\ \LL^{(1)}(v\xi^2)^{{\rm t}_1}\ \Der_1^{-2}=
\Der_2^{2}\  \LL^{(2)}(v\xi^2)^{{\rm t}_2}\ \Der_2^{-2}\ \Big(\LL^{(2)}(v)^{-1}\Big)^{{\rm t}_2}
\end{equation*} 
 which proves equality in \r{Cen1}. To prove centrality of the elements $Z^\pm(v)$ 
we consider the chain of equalities 
\begin{equation*}
\begin{split}
&Z(u)\Id_1\ \LL^{(2)}(v)=\Der^2_1\ \LL^{(1)}(u\xi^2)^{{\rm t}_1}\ \Der^{-2}_1\Big(\LL^{(1)}(u)^{-1}\Big)^{{\rm t}_1}
\LL^{(2)}(v)=\\
&\quad= \Der^2_1\ \LL^{(1)}(u\xi^2)^{{\rm t}_1}\ \Der^{-2}_1
\Big(\RR_{21}(v,u)^{{\rm t}_1}\Big)^{-1}\LL^{(2)}(v)\ \Big(\LL^{(1)}(u)^{-1}\Big)^{{\rm t}_1}
\RR_{21}(v,u)^{{\rm t}_1}=\\
&\quad=x(u,v) \Der_1^2\ 
\LL^{(1)}(u\xi^2)^{{\rm t}_1} \RR_{12}(u\xi^2,v)^{{\rm t}_1}
\LL^{(2)}(v)\ \Der_1^{-2}\Big(\LL^{(1)}(u)^{-1}\Big)^{{\rm t}_1}
\RR_{21}(v,u)^{{\rm t}_1}=\\
&\quad=x(u,v) \LL^{(2)}(v)\ \Der_1^2\ \RR_{12}(u\xi^2,v)^{{\rm t}_1}
\LL^{(1)}(u\xi^2)^{{\rm t}_1} \Der_1^{-2}\Big(\LL^{(1)}(u)^{-1}\Big)^{{\rm t}_1}
\RR_{21}(v,u)^{{\rm t}_1}=\\
&\quad=\LL^{(2)}(v)\ \Big(\RR_{21}(v,u)^{{\rm t}_1}\Big)^{-1}\ Z(u)\Id_1\ 
\RR_{21}(v,u)^{{\rm t}_1}= \LL^{(2)}(v)\ Z(u)\Id_1\,,
\end{split}
\end{equation*}
where  $x(u,v)$ is defined by \r{xuv}.
For these calculations  one has to use $\RR\LL\LL$ commutation relations and 
equalities  \r{crossingA} and \r{Cross} for $\RR$-matrices \r{R-matA} and 
\r{R-mat}. \qed

\begin{remark}\label{qdet}
Existence of the central element $Z(u)$ for the Yangian  $Y(\mathfrak{gl}_N)$ 
was mentioned in  \cite{Naz90}. In this paper 
   a {\it quantum Liouville formula} for the Yangian 
 was considered. Analogous relation in the case of the algebra 
 $U_q(A^{(1)}_{N-1})$ takes the form 
\begin{equation}\label{RA9}
Z^\pm(v)=  \prod_{s=1}^{N} 
\frac{k^\pm_s (v q^{-2 s})}{k^\pm_s (v q^{-2(s-1)})}  = \frac{\mbox{q-det}\,
\big(\LL^\pm(vq^{-2})\big)}{\mbox{q-det}\,\big(\LL^\pm(v)\big)}
\end{equation}
and can be proved in the same way as in the Yangian case \cite{Mol02}.
In \r{RA9} $k^\pm_\ell(v)$ are diagonal Gauss coordinates introduced by \r{Gauss1}.
\end{remark}

We set the central elements $Z^\pm(u)$ 
equal to 1 in the algebra $U_q(\gaf)$. We denote by 
$\tilde U_q(A^{(1)}_{N-1})$ the algebra defined by $U_q(\mathfrak{gl}_N)$-invariant 
$\RR$-matrix \r{R-matA} without any restrictions to these central elements.

The pole structure of $\RR$-matrix for $\gaf=B^{(1)}_n,C^{(1)}_n,D^{(1)}_n,A^{(2)}_{N-1}$
given by \r{poles} yields other central elements in the corresponding algebras $U_q(\gaf)$. 
We have following 
\begin{prop} 
There are central elements $z^\pm(v)\in U_q^\pm(\gaf)$ for 
$\gaf=B^{(1)}_n$, $C^{(1)}_n$, $D^{(1)}_n$  and $A^{(2)}_{N-1}$ given by the equalities 
\begin{equation}\label{cent5}
z^\pm(v)\ \Id=
\Der\ \LL^\pm(v\xi)^{{\rm t}}\ \Der^{-1}\ \LL^\pm(v)=
\LL^\pm(v)\ \Der\ \LL^\pm(v\xi)^{{\rm t}}\ \Der^{-1}.
\end{equation}
Again, \eqref{cent5} means that product of the matrices 
\begin{equation*}
\Der\ \LL^\pm(v\xi)^{{\rm t}}\ \Der^{-1}\ \LL^\pm(v)\quad\mbox{and}\quad 
\LL^\pm(v)\ \Der\ \LL^\pm(v\xi)^{{\rm t}}\ \Der^{-1}
\end{equation*} 
are proportional 
to the unity operator $\Id$ and the proportionality coefficients are central elements.
They are related to  $Z^\pm(v)$ by the relations
\begin{equation}\label{cen7}
Z^\pm(v)=z^\pm(v\xi)\ z^\pm(v)^{-1}.
\end{equation}
\end{prop}

{\it Proof.}  Calculating residue at $u=v\xi$ in the  commutation relation \r{RLL}  one gets 
\begin{equation*}\label{cent1}
\Der_1^{-1}\ \Per_{12}^{{\rm t}_1}\ \Der_1\ \LL^{(1)}(v\xi)\ \LL^{(2)}(v)=
\LL^{(2)}(v)\ \LL^{(1)}(v\xi)\ \Der_1^{-1}\ \Per_{12}^{{\rm t}_1}\ \Der_1\,,
\end{equation*}
which is equivalent to 
\begin{equation*}\label{cent4}
 \Der_1\ \LL^{(1)}(v\xi)^{{\rm t}_1}\ \Der_1^{-1}\ \LL^{(1)}(v)=
\LL^{(2)}(v)\ \Der_2\ \LL^{(2)}(v\xi)^{{\rm t}_2}\ \Der_2^{-1}.
\end{equation*}
This  proves \r{cent5}.
 
To prove that the elements $z^\pm(u)$ are central elements in the algebra $U_q(\gaf)$
we consider the product $z(u)\Id_1\ \LL^{(2)}(v)$
and chain of equalities 
\begin{equation*}
\begin{split}
&z(u)\Id_1\ \LL^{(2)}(v)=\Der_1\ 
\LL^{(1)}(u\xi)^{{\rm t}_1}\ \Der_1^{-1}\ \LL^{(1)}(u)
\LL^{(2)}(v)=\\
&=\Der_1\ \LL^{(1)}(u\xi)^{{\rm t}_1}\ \Der_1^{-1}\ \RR_{12}(u,v)^{-1}
\ \LL^{(2)}(v)\ \LL^{(1)}(u)\ \RR_{12}(u,v)=\\
&=x(u,v\xi)\ \Der_1\ \LL^{(1)}(\xi u)^{{\rm t}_1}\ 
\RR_{12}(u\xi, v)^{{\rm t}_1}\ \LL^{(2)}(v)\ \Der_1^{-1}\ \LL^{(1)}(u)\ 
\RR_{12}(u,v)=\\
&=x(u,v\xi)\ \Der_1\ \LL^{(2)}(v)\
\RR_{12}(u\xi, v)^{{\rm t}_1}\ \LL^{(1)}(u\xi)^{{\rm t}_1}\ 
 \Der_1^{-1}\ \LL^{(1)}(u)\ 
\RR_{12}(u,v)=\\
&=\LL^{(2)}(v)\
\RR_{12}(u,v)^{-1}\ \Der_1\ \LL^{(1)}(\xi u)^{{\rm t}_1}\ 
 \Der_1^{-1}\ \LL^{(1)}(u)\ 
\RR_{12}(u,v)=\\
&=\LL^{(2)}(v)\ \RR_{12}(u,v)^{-1}\ z(u)\Id_1\  \RR_{12}(u,v)=\LL^{(2)}(v)\ z(u)\Id_1.
\end{split}
\end{equation*}
 Equality \r{cen7} can be proved by expressing 
$\Big(\LL^\pm(v)^{-1}\Big)^{{\rm t}}$ from \r{cent5} and substituting it into \r{Cen1}.
\qed

For the algebras $U_q(\gaf)$ with $\gaf=B^{(1)}_n,C^{(1)}_n,D^{(1)}_n,A^{(2)}_{N-1}$
we set central elements $z^\pm(v)=1$. Then 
 equalities \r{cent5} take the form 
\begin{equation*}\label{ident}
\Der\ \LL^\pm(v\xi)^{{\rm t}}\ \Der^{-1}= \LL^\pm(v)^{-1}\,,
\end{equation*} 
or
\begin{equation}\label{identalt}
\Der\ \hLL^\pm(v\xi)\ \Der^{-1}= \LL^\pm(v)\,,
\end{equation} 
where transposed-inversed $\LL$-operators $\hLL^\pm(u)$ are defined as
\begin{equation}\label{hLL}
\hLL^\pm(u)=\Big(\LL^\pm(u)^{{\rm t}}\Big)^{-1}\,.
\end{equation}
Due to \r{cen7} the central elements $Z^\pm(v)$ also equal to 1 when $z^\pm(v)=1$.
Then equality \r{Cen1} can be written in the form 
\begin{equation}\label{cen8}
\hLL^\pm(v)=\Big(\LL^\pm(u)^{{\rm t}}\Big)^{-1}
= \Der^{-2}\ \Big(\LL^\pm(v\xi^{-2})^{-1}\Big)^{{\rm t}}\ \Der^{2}
\end{equation}
and describes the relations between order of transposition and taking inverse of 
the fundamental $\LL$-operators in the algebra $U_q(\gaf)$.

One can check  
that  $\LL$-operators $\hLL^\pm(u)$ given by \r{hLL} 
satisfies the same commutation relations  \r{RLL}.  Let us  
apply to \r{RLL} the transposition \r{trans} in both auxiliary spaces and use \r{reflec}
to get
\begin{equation*}
\LL^{(1)}(u)^{{\rm t}_1}\  \LL^{(2)}(v)^{{\rm t}_2}\ \RR_{12}(u,v)
 =
 \RR_{12}(u,v)\  \LL^{(2)}(v)^{{\rm t}_2}\  \LL^{(1)}(u)^{{\rm t}_1}\,.
\end{equation*}
Multiplying from both sides of this equality first by $\hLL^{(1)}(u)$ and then by 
 $\hLL^{(2)}(v)$ one gets 
\begin{equation*}\label{hRLL}
\RR_{12}(u,v)\  \hLL^{(1)}(u)\ \hLL^{(2)}(v) =
\hLL^{(2)}(v)\ \hLL^{(1)}(u)\  \RR_{12}(u,v)\,.
\end{equation*}
Summarizing we conclude   that the map 
\begin{equation*}\label{autoL}
\LL^\pm(u)\to \hLL^\pm(u)
\end{equation*}
moves the algebra $U_q(\gaf)$ into the  algebra 
given by the same commutation relation \r{RLL} but for the 
transposed-inversed $\LL$-operators $\hLL^\pm(u)$. 
One can also check that central elements $\hat Z^\pm(v)$ and $\hat z^\pm(v)$ 
defined by \r{Cen1} and \r{cent5} with $\LL^\pm(v)$ replaced by $\hLL^\pm(v)$
are related to the central elements $Z^\pm(v)$ and $z^\pm(v)$ as follows
\begin{equation*}
\hat Z^\pm(v)=Z^\pm(v)^{-1},\qquad \hat z^\pm(v)=z^\pm(v)^{-1}.
\end{equation*}

\section{Gauss coordinates} 
\label{GC}

It is known  \cite{KhT93} that Gauss coordinates of $\LL$-operators introduced below by 
the equality \r{Gauss1} are related to the Cartan-Weyl generators 
of the corresponding to $U_q(\gaf)$ quantum  affine  algebras 
  $U_q(\widehat{\mathfrak{g}})$. 
The Cartan-Weyl generators satisfy certain ordering properties described 
in details in \cite{EKhP07} and shortly presented in the section~\ref{nor-ord}. 
In this paper we consider 
Gauss decomposition of the fundamental $\LL$-operators of the algebra $U_q(\gaf)$
\begin{equation}\label{Gauss1}
\LL^\pm_{i,j}(u)=\sum_{\ell\leq{\rm min}(i,j)} \FF^\pm_{j,\ell}(u)\ k^\pm_\ell(u)\ \EE^\pm_{\ell,i}(u)\,,
\end{equation}
where one assumes that  $\FF^\pm_{i,i}(u)=\EE^\pm_{i,i}(u)=1$ for 
$1\leq i\leq N$.

Gauss decompositions formula for the matrix entries of $\LL$-operators is  associated with the products of lower triangular, diagonal and upper triangular matrices 
\begin{equation}\label{Gauss3}
\begin{split}
\LL^\pm(u)^{\rm t}&= \sum_{1\leq i,j\leq N} \LL^\pm_{i,j}(u)\ \E^{\rm t}_{ij}=\\
&=\sk{\sum_{1\leq i<j\leq N} \E_{ij}^{\rm t}\ \FF^\pm_{j,i}(u)}\cdot
\sk{\sum_{1\leq i\leq N} \E_{ii}^{\rm t}\ k^\pm_{i}(u)}\cdot
\sk{\sum_{1\leq i<j\leq N} \E_{ji}^{\rm t}\ \EE^\pm_{i,j}(u)}\,.
\end{split}
\end{equation}

Equality \r{Gauss3} allows to obtain Gauss decomposition 
of $\LL$-operators $\hLL^\pm(u)$. Indeed, 
using multiplication rule for $\E^{\rm t}_{ij}=\varepsilon_i\varepsilon_j\ \E_{j'i'}$
\begin{equation*}\label{muru}
\E^{\rm t}_{ij}\cdot \E^{\rm t}_{kl}=\delta_{il}\ \E^{\rm t}_{kj}
\end{equation*}
and taking inverse of both sides of the equality  \r{Gauss3}  
\begin{equation}\label{Gau33}
\begin{split}
\hLL^\pm(u)&= \Big(\big(\LL^\pm(u)\big)^{\rm t}\Big)^{-1}
= \sum_{1\leq i,j\leq N} \E_{ij}\ \hLL^\pm_{i,j}(u)=\\
&=\sk{\sum_{1\leq i\leq j\leq N} \E_{ji}^{\rm t}\ \tEE^\pm_{i,j}(u)}\cdot
\sk{\sum_{1\leq i\leq N} \E_{ii}^{\rm t}\ k^\pm_{i}(u)^{-1}}\cdot
\sk{\sum_{1\leq i\leq j\leq N} \E_{ij}^{\rm t}\ \tFF^\pm_{j,i}(u)}
\end{split}
\end{equation} 
one obtains 
 \begin{equation}\label{GChL}
 \hLL^\pm_{i,j}(u)=\varepsilon_i \varepsilon_j 
 \sum_{ \ell\leq {\rm min}(i,j)}\tEE^\pm_{i',\ell'}(u)\ k^\pm_{\ell'}(u)^{-1}\
 \tFF^\pm_{\ell',j'}(u)\,,
 \end{equation}
where equality $\varepsilon_i\varepsilon_j=\varepsilon_{i'}\varepsilon_{j'}$ 
was used. 
Gauss coordinates $\tFF^\pm_{j,i}(u)$ and $\tEE^\pm_{i,j}(u)$ 
in \r{Gau33} and \r{GChL} satisfy recurrence relations 
\begin{equation*}
\sum_{i\leq\ell\leq j}\FF^\pm_{j,\ell}(u)\tFF^\pm_{\ell,i}(u)= \delta_{ij}\quad
\mbox{and}\quad 
\sum_{i\leq\ell\leq j}\EE^\pm_{i,\ell}(u)\tEE^\pm_{\ell,j}(u)=\delta_{ij}
\end{equation*}
which can be resolved in the form 
\begin{equation*}\label{tFF}
\tFF^\pm_{j,i}(u) =  -\FF^\pm_{j,i}(u)+\sum_{\ell=1}^{j-i-1}(-)^{\ell+1}
\sum_{j>i_\ell>\cdots>i_1>i} \FF^\pm_{j,i_\ell}(u)  \FF^\pm_{i_\ell,i_{\ell-1}}(u)\cdots 
\FF^\pm_{i_2,i_1}(u)  \FF^\pm_{ i_1,i}(u)
\end{equation*}
and 
\begin{equation*}\label{tEE}
\tEE^\pm_{i,j}(u) =  -\EE^\pm_{i,j}(u)+\sum_{\ell=1}^{j-i-1}(-)^{\ell+1}
\sum_{j>i_\ell>\cdots>i_1>i} \EE^\pm_{i,i_1 }(u)  \EE^\pm_{i_1,i_2}(u)\cdots 
\EE^\pm_{i_{\ell-1},i_\ell}(u)  \EE^\pm_{i_\ell,j}(u)\,.
\end{equation*}

\subsection{Normal ordering of the Gauss coordinates}\label{nor-ord}

 In order to obtain commutation relations for Gauss coordinates from  \r{TM-1} 
 one can use  the normal ordering of the Cartan-Weyl generators  \cite{EKhP07}. 

Let $U^\pm_f$, $U^\pm_e$ and $U^\pm_k$ be subalgebras of $U_q(\gaf)$
formed by the modes of the Gauss coordinates $\FF^\pm_{j,i}(u)$, $\EE^\pm_{i,j}(u)$
and $k^\pm_j(u)$.  The fact that these unions of generators are 
subalgebras follows from the identification of modes of Gauss coordinates with 
Cartan-Weyl generators \cite{KhT93}.
It is known that Cartan-Weyl generators 
have two natural circular orderings  which imply the normal ordering of the subalgebras 
formed by the Gauss coordinates. These orderings are  
\begin{equation}\label{order1}
\cdots\prec U^-_k\prec U^-_f\prec U^+_f\prec U_k^+\prec U^+_e\prec U^-_e\prec U^-_k\prec\cdots
\end{equation}
or 
\begin{equation}\label{order2}
\cdots\prec U^+_k\prec U^+_f\prec U^-_f\prec U_k^-\prec U^-_e\prec U^+_e\prec U^+_k\prec\cdots
\end{equation}

If one places subalgebras $U^\pm_f$, $U^\pm_e$ and $U^\pm_k$
onto circles
\begin{equation}\label{circ}
\begin{array}{ccccc}
&U^-_e&&U^+_e& \\[3mm]
U^-_k& &\circlearrowleft& & U^+_k\\[3mm]
&U^-_f&&U^+_f&
\end{array}
\qquad\mbox{and} \qquad
\begin{array}{ccccc}
&U^-_e&&U^+_e& \\[3mm]
U^-_k& &\circlearrowright& & U^+_k\\[3mm]
&U^-_f&&U^+_f&
\end{array}
\end{equation}
then ordering \r{order1} is counterclockwise  in the left circle  
and the ordering \r{order2} is  clockwise in the right circle of \r{circ}. 
The general theory of the Cartan-Weyl basis allows to prove that in both types 
of ordering the unions of subalgebras $U^\pm_f$, $U^\pm_e$ and $U^\pm_k$
along smallest arcs between starting and ending points are 
subalgebras in  $U_q(\gaf)$. 
For example, the union of subalgebras $U^+_f\cup U^+_k$ 
or $U^-_f\cup U^+_f\cup U^+_k$ or  $U_q^+(\gaf)=U^+_f\cup U^+_k\cup U^+_e$ 
and so on are subalgebras in  $U_q(\gaf)$.

The notion of the normal ordering yields a powerful practical tool to 
get relations for the Gauss coordinates of the specific type. In any relation 
which contains Gauss coordinates of the different types one first has to 
order all monomials according to \r{order1} or \r{order2}  and then single out 
all the terms which belong to the one of subalgebras which is composed from the 
Gauss coordinates of the necessary type. 
We call this procedure {\it a restriction} to subalgebras in $U_q(\gaf)$ and will 
use this method to get relations between Gauss coordinates from $\RR\LL\LL$-commutation 
relations \r{TM-1}.

Subalgebras $U^\pm_q(\gaf)=U^\pm_f\cup U^\pm_k\cup U^\pm_e$
 were already introduced above as  Borel subalgebras in $U_q(\gaf)$. 
To descride so called 'new realization' of these algebras in terms of the currents \cite{D88}
one has to consider different type Borel subalgebras $U_f=U^-_f\cup U^+_f\cup U^+_k$
and $U_e=U^+_e\cup U^-_e\cup U^-_k$.
In \cite{DKhP00-1} certain   projections $\Pfpm$  and $\Pepm$ 
onto intersections of the Borel subalgebras of the different types were introduced. 
These projections were further investigated in \cite{EKhP07} for the ordering 
\r{order1} and was used for the first time
in  \cite{KhP05} to describe the off-shell Bethe vectors 
or weight functions in terms of the Cartan-Weyl generators. One can check that 
 the action of the projections $\Pfpm$  and $\Pepm$ 
onto Borel subalgebras $U_f$ and $U_e$ introduced in \cite{DKhP00-1} 
coincides with restrictions onto  subalgebras
$U_f^\pm$ and $U_e^\pm$  defined for the ordering \r{order1}.

\section{Embedding theorem}\label{embed}

Each algebra $\gaf$ of the type $B^{(1)}_{n}$, $C^{(1)}_{n}$, $D^{(1)}_{n}$ and 
$A^{(2)}_{N-1}$  has rank $n$ as rank of the 
underlying finite dimensional algebra. To stress this fact 
we will use notation $U_q^n(\gaf)$ to denote explicitly rank 
for any of the quantum loop algebras considered in this paper. 
Following ideas of the paper \cite{JLM19} 
 we consider in this section embedding of smaller algebras 
  $U_q^{n-1}(\gaf)\hookrightarrow U_q^n(\gaf)$. 
To note that  $\RR$-matrix corresponds to the algebra $U_q^n(\gaf)$
we will use superscript $\RR^n(u,v)$, $\AR^n(u,v)$, $\QR^n(u,v)$, etc.

In this paper we use Gauss decomposition 
of the $\LL$-operators  
for the algebra  $U_q^n(\gaf)$
given by \r{Gauss1}  
\begin{equation}\label{Em1}
\begin{split}
\LL^\pm_{i,j}(u)&=\sum_{1\leq 
\ell\leq{\rm min}(i,j)} \FF^\pm_{j,\ell}(u)\ k^\pm_\ell(u)\ \EE^\pm_{\ell,i}(u)=\\
&=\MM^\pm_{i,j}(u)+\FF^\pm_{j,1}(u)k^\pm_1(u)\EE^\pm_{1,i}(u)
=\MM^\pm_{i,j}(u)+\LL^\pm_{1,j}(u)\LL^\pm_{1,1}(u)^{-1}\LL^\pm_{i,1}(u)\,.
\end{split}
\end{equation}
Let us consider matrix entries $\MM^\pm_{i,j}(u)$ defined by \r{Em1} for $1< i,j < N$. 
These are matrix entries of the 
$(N-2)\times (N-2)$ matrix $\MM^\pm(u)$ of the fundamental $\LL$-operators 
for the algebra $U_q^{n-1}(\gaf)$.  For $1< i,j < N$
matrix entries $\MM^\pm_{i,j}(u)$ have Gauss decomposition 
\begin{equation}\label{Em2}
\MM^\pm_{i,j}(u)=\sum_{2\leq 
\ell\leq{\rm min}(i,j)} \FF^\pm_{j,\ell}(u)\ k^\pm_\ell(u)\ \EE^\pm_{\ell,i}(u)\,.
\end{equation}
 For $\gaf=B^{(1)}_n,C^{(1)}_n,D^{(1)}_n$ and $A^{(2)}_{N-1}$
we have following 
\begin{thm}\label{thm-emb}
The commutation relations for the $U_q^{n-1}(\gaf)$ matrix entries 
$\MM^\pm_{i,j}(u)$ follow from the Yang-Baxter equation \r{YB} and 
the commutation relations \r{RLL} in $U_q^{n}(\gaf)$ 
and take the form ($\mu,\nu=\pm$)
\begin{equation}\label{Em20}
\RR_{12}^{n-1}(u,v) \ (\MM^\mu(u)\ot\Id) \, (\Id\ot\MM^\nu(v))=
(\Id\ot\MM^\nu(v))\, (\MM^\mu(u)\ot\Id)\ \RR_{12}^{n-1}(u,v)\,.
\end{equation}
\end{thm}

To prove this theorem we  formulate auxiliarly lemmas~\ref{lem51} and \ref{lem52}.
Let $\Ll^{(1,2)}(u)$ be fused
  $\LL$-operator defined as (we again skip superscripts of $\LL$-operators to avoid bulky 
  notations)
\begin{equation*}\label{Em6}
\Ll^{(1,2)}(u)=\RR(1,q^2)\ \LL^{(1)}(u)\ \LL^{(2)}(q^2u)=
\LL^{(2)}(q^2u)\ \LL^{(1)}(u)\ \RR(1,q^2)\,.
\end{equation*}
One can calculate its $(i,j;1,1)$ matrix element
\begin{equation}\label{Em7}
\begin{split}
\Ll_{i,j;1,1}(u)&=\<i,1|\Ll^{(1,2)}(u)|j,1\>=
\sum_{k,l=1}^N\RR_{i,k;1,l}(1,q^2)\LL_{k,j}(u)\LL_{l,1}(q^2u)=\\
&=\LL_{i,j}(u)\LL_{1,1}(q^2u)-q\LL_{1,j}(u)\LL_{i,1}(q^2u)\,,
\end{split}
\end{equation}
where $|i,j\>=|i\>\ot|j\>$ and 
$\<i,j|=\<i|\ot\<j|$ are vectors in $\sk{\CC^N}^{\ot2}$ such that $\<i|j\>=\delta_{ij}$. 

The commutation relations for $1<i<N$
\begin{equation*}\label{Em4}
\LL_{1,1}(u)^{-1}\LL_{i,1}(u)=q\,\LL_{i,1}(q^2u)\LL_{1,1}(q^2u)^{-1}
\end{equation*}
and \r{Em7} implies
that $\LL$-operators $\MM(u)$ for the algebra  
$U_q^{n-1}(\gaf)$  can be presented as 
\begin{equation}\label{Em8}
\MM_{i,j}(u)=\Ll_{i,j;1,1}(u)\ \LL_{1,1}(q^2u)^{-1}\,,
\end{equation}
where $1<i,j<N$.

\begin{lemma}\label{lem51}
There is a commutativity of the matrix entries in $U_q^{n}(\gaf)$
\begin{equation*}\label{Em144}
\LL_{1,1}(u)\  \MM_{i,j}(v)= \MM_{i,j}(v)\ \LL_{1,1}(u),\quad 1<i,j<N\,.
\end{equation*}
\end{lemma}

According to \r{Em8} matrix entries $\MM_{i,j}(u)$ are proportional to 
the matrix entries $\Ll_{i,j;1,1}(u)$ up to commuting with them invertible 
operator $\LL_{1,1}(v)$.
It yields that the commutation relations for $\MM_{i,j}(u)$ should coincide 
with the commutation relations of $\Ll_{i,j;1,1}(u)$. 
To find the commutation relations for the matrix entries $\Ll_{i,j;1,1}(u)$
we need following
\begin{lemma}\label{lem52}
 There are equalities for $1<i,j<N$
\begin{equation}\label{Em18}
\begin{split}
&\RR^n_{12}(1,q^2)\RR^n_{34}(1,q^2)\RR^n_{14}(u,q^2v)\RR^n_{13}(u,v)|i,1,j,1\>=\\
&\quad=\RR^n_{12}(1,q^2)\RR^n_{34}(1,q^2)\RR^{n-1}_{13}(u,v)|i,1,j,1\>
\end{split}
\end{equation}
and 
\begin{equation}\label{Em19}
\begin{split}
&\<i,1,j,1|\RR^n_{13}(u,v)\RR^n_{14}(u,q^2v)\RR^n_{34}(1,q^2)\RR^n_{12}(1,q^2)=\\
&\quad=\<i,1,j,1|\RR^{n-1}_{13}(u,v)\RR^n_{34}(1,q^2)\RR^n_{12}(1,q^2)
\end{split}
\end{equation}
where $|i,k,j,l\>=|i\>\ot|k\>\ot|j\>\ot|l\>$ and 
$\<i,k,j,l|=\<i|\ot\<k|\ot\<j|\ot\<l|$ are vectors in $\sk{\CC^N}^{\ot4}$. 
\end{lemma}

Proofs of the lemmas~\ref{lem51} and \ref{lem52} can be found in appendix~\ref{prf-lem}.

To prove theorem~\ref{thm-emb} we consider 
$\RR\LL\LL$-commutation relations for $\LL$-operators $\Ll(u)$ and $\Ll(v)$ 
\r{Em11} and   for $1<i_1,j_1,i_2,j_2<N$ take the matrix element of this commutation 
relation 
\begin{equation}\label{Em21}
\begin{split}
&\<i_1,1,j_1,1|
\RR^n_{23}(q^2u,v)\RR^n_{13}(u,v)\RR^n_{24}(u,v)\RR^n_{14}(u,q^2v)
\RR^n_{12}(1,q^2) \RR^n_{34}(1,q^2)\times\\
&\qquad \times \LL^{(1)}(u)\LL^{(2)}(q^2u) \LL^{(3)}(v)\LL^{(4)}(q^2v)
|i_2,1,j_2,1\>
=\\
&\quad =\<i_1,1,j_1,1|\LL^{(4)}(q^2v)\LL^{(3)}(v)\LL^{(2)}(q^2u)\LL^{(1)}(u)\times\\
&\qquad \times \RR^n_{34}(1,q^2) \RR^n_{12}(1,q^2) 
 \RR^n_{14}(u,q^2v)\RR^n_{24}(u,v)\RR^n_{13}(u,v)\RR^n_{23}(q^2u,v)|i_2,1,j_2,1\>
\end{split}
\end{equation}
Let us transform last line in \r{Em21} using lemma~\ref{lem52}, equality~\r{Em15}  and Yang-Baxter 
equation \r{YB}. 
We have 
\begin{equation}\label{Em22}
\begin{split}
&\RR^n_{34}(1,q^2) \RR^n_{12}(1,q^2) 
 \RR^n_{14}(u,q^2v)\RR^n_{24}(u,v)\RR^n_{13}(u,v)\RR^n_{23}(q^2u,v)|i_2,1,j_2,1\>=\\
 &\quad= \RR^n_{12}(1,q^2)\RR^n_{13}(u,v) \RR^n_{14}(u,q^2v)
 \RR^n_{34}(1,q^2)\RR^n_{24}(u,v)\RR^n_{23}(q^2u,v)|i_2,1,j_2,1\>=\\
 &\quad=f(q^2u,v)\RR^n_{12}(1,q^2) \RR^n_{34}(1,q^2)
 \RR^n_{14}(u,q^2v)\RR^n_{13}(u,v)|i_2,1,j_2,1\>=\\
&\quad=f(q^2u,v)\RR^n_{12}(1,q^2) \RR^n_{34}(1,q^2)\RR^{n-1}_{13}(u,v)|i_2,1,j_2,1\>\,.
\end{split}
\end{equation}
At the second step of this calculation we used equality \r{Em15} taken at $u\to q^2u$ 
and scaling invariance of $\RR$-matrix \r{scaling}. 

Analogously first line in \r{Em21} can be transformed to 
\begin{equation}\label{Em23}
\begin{split}
&\<i_1,1,j_1,1|
\RR^n_{23}(q^2u,v)\RR^n_{13}(u,v)\RR^n_{24}(u,v)\RR^n_{14}(u,q^2v)
\RR^n_{12}(1,q^2) \RR^n_{34}(1,q^2)=\\
&\quad = f(q^2u,v) \<i_1,1,j_1,1|\RR^{n-1}_{13}(u,v)\RR^n_{12}(1,q^2) \RR^n_{34}(1,q^2)\,,
\end{split}
\end{equation}
where we used \r{Em16} at $v\to q^{-2}v$.

Equalities \r{Em22} and \r{Em23} allow to rewrite \r{Em21} in the form
\begin{equation*}\label{Em24}
\begin{split}
&\<i_1,1,j_1,1|\RR^{n-1}_{13}(u,v)\ \Ll^{(1,2)}(u)\Ll^{(3,4)}(v)|i_2,1,j_2,1\>=\\
&\qquad = \<i_1,1,j_1,1|\Ll^{(3,4)}(v)\Ll^{(1,2)}(u)\ \RR^{n-1}_{13}(u,v)|i_2,1,j_2,1\>
\end{split}
\end{equation*}
which proves the statement of  theorem \r{Em20} due to lemma~\ref{lem51}
and relation \r{Em8}. \qed

Theorem~\ref{thm-emb} implies that 
in order to find the commutation relations between Gauss coordinates in the  
algebra $U_q^{n}(\gaf)$    
it is sufficient to obtain these commutation relations for the smallest rank
nontrivial algebras. We will find such commutation relations 
 in the algebras $U_q(\gaf)$
for $\gaf$ of the types
$B^{(1)}_n$, $C^{(1)}_n$, $D^{(1)}_n$, $A^{(2)}_{N-1}$ in appendix~\ref{smrk}.

We can formulate analogous  statement  
 for the algebra $U_q(A^{(1)}_{N-1})$. Let $\MM_{i,j}(u)$ for 
$1<i,j\leq N$
be matrix entries of the $\LL$-operators for the algebra 
$U_q(A^{(1)}_{N-2})$ defined by \r{Em2}. Denote by 
$\AR^N(u,v)$ $\RR$-matrix \r{R-matA} for the algebra $U_q(A^{(1)}_{N-1})$.
Using similar arguments as above 
we can prove following 
\begin{prop}\label{pr-em-A}
The commutation relations of the matrix entries $\MM_{i,j}(u)$ and their 
Gauss coordinates of the fundamental 
$\LL$-operators of the algebra $U_q(A^{(1)}_{N-2})$ for $1<i,j\leq N$
follow from the commutation relations \r{RLL} for the algebra 
 $U_q(A^{(1)}_{N-1})$ and take the same form with $\RR$-matrix 
 $\AR^{N-1}(u,v)$.
\end{prop}

We are not going to provide a proof of this proposition since it can be performed 
in a similar way as the proof of theorem~\ref{thm-emb}. Practical meaning 
of this proposition is that in order to obtain the commutation relations between 
Gauss coordinates for the algebra $U_q(A^{(1)}_{N-1})$ it is sufficient 
to consider the commutation relations for the algebras at small values of $N$.  

\subsection{Embedding in terms of the Gauss coordinates}

Let us introduce 'alternative' to \r{Gauss1} 
Gauss decomposition of the fundamental $\LL$-operators
\begin{equation*}\label{Gau333}
\begin{split}
\LL(u)=&\sk{\sum_{1\leq i\leq j\leq N} \E_{ji}\ \bEE_{i,j}(q^{-2(i-1)}u)}\times \\
&\qquad\times \sk{\sum_{1\leq i\leq N} \E_{ii}\ \bk_{i}(q^{-2(i-1)}u)}\cdot
\sk{\sum_{1\leq i\leq j\leq N} \E_{ij}\ \bFF_{j,i}(q^{-2(i-1)}u)}\,,
\end{split}
\end{equation*} 
where shifts by $q^{-2(i-1)}$ in the arguments of  'alternative' Gauss coordinates 
$\bFF_{j,i}(u)$, $\bEE_{i,j}(u)$ and $\bk_i(u)$ are introduced for the further 
convenience. In terms of these Gauss coordinates matrix entries of $\LL$-operators 
have the form 
\begin{equation}\label{Em333}
\LL_{i,j}(u)=\sum_{1\leq 
\ell\leq{\rm min}(i,j)} \bEE_{\ell,i}(q^{-2(\ell-1)}u)\ \bk_\ell(q^{-2(\ell-1)}u)\ 
\bFF_{j,\ell}(q^{-2(\ell-1)}u)\,.
\end{equation}
 In  \r{Em333} we assume that   $\bFF^\pm_{i,i}(u)=\bEE^\pm_{i,i}(u)=1$ for 
$1\leq i\leq N$.

Our goal is to find relations between  Gauss coordinates $\FF_{j,i}(u)$, $\EE_{i,j}(u)$, $k_j(u)$ 
and $\bFF_{j,i}(u)$, $\bEE_{i,j}(u)$, $\bk_j(u)$. This is given by 
\begin{prop}\label{inverse}
For $\gaf=B^{(1)}_n,C^{(1)}_n,D^{(1)}_n,A^{(2)}_{N-1}$ 
Gauss coordinates $\bFF^\pm_{j,i}(u)$, $\bEE^\pm_{i,j}(u)$ and $\bk^\pm_j(u)$ are related 
to the initial Gauss coordinates $\FF^\pm_{j,i}(u)$, $\EE^\pm_{i,j}(u)$ and $k^\pm_j(u)$:
\begin{equation}\label{Finv}
\bFF^\pm_{j,i}(u)=q\FF^\pm_{j,i}(q^{-2}u)\,,
\end{equation}
\begin{equation}\label{Einv}
\bEE^\pm_{i,j}(u)=q^{-1}\EE^\pm_{i,j}(q^{-2}u)\,,
\end{equation}
\begin{equation}\label{Kinv}
\bk^\pm_{\ell}(u)=k^\pm_{\ell}(u)
\prod_{s=1}^{\ell-1}\frac{k^\pm_s(q^{2(\ell-s)}u)}{k^\pm_s(q^{2(\ell-s-1)}u)}\,,
\end{equation}
where $i<j<i'$, $1\leq i\leq n$, $1\leq\ell\leq n+1$ for odd $N=2n+1$  and 
$1\leq i\leq n-1$, $1\leq\ell\leq n$ for $N=2n$ even. 
Moreover, for $1\leq i<j\leq N$ and $1\leq\ell\leq N$
\begin{equation}\label{Finvall}
\bFF^\pm_{j,i}(u)=\varepsilon_i\varepsilon_j\ q^{\bar\imath-\bar\jmath}\
\tFF^\pm_{i',j'}(q^{2(i-1)}\xi u)\,,
\end{equation}
\begin{equation}\label{Einvall}
\bEE^\pm_{i,j}(u)=\varepsilon_i\varepsilon_j\ q^{\bar\jmath-\bar\imath}\
\tEE^\pm_{j',i'}(q^{2(i-1)}\xi u)\,,
\end{equation}
\begin{equation}\label{Kinvall}
\bk^\pm_{\ell}(u)=k^\pm_{\ell'}(q^{2(\ell-1)}\xi u)^{-1}\,.
\end{equation}
\end{prop}

The proof of this proposition is given in appendix~\ref{appB}.

Note that equality \r{Kinv} can be written in the form 
\begin{equation*}\label{Em50}
\bk^\pm_{\ell}(u)=k^\pm_{\ell}(u)\frac{\bk^\pm_{\ell-1}(q^2u)}{k^\pm_{\ell-1}(u)}
\end{equation*}
which is consequence of the embedding relation \r{Em43} at each step of the 
embedding. Moreover, we can exclude $\bk^\pm_\ell(u)$ from \r{Kinv} and \r{Kinvall}
to obtain 
\begin{equation}\label{Ksol}
k^\pm_{\ell'}(u)=k^\pm_{\ell}(q^{-2(\ell-1)}\xi^{-1}u)^{-1}
\prod_{s=1}^{\ell-1}\frac{k^\pm_s(q^{-2s}\xi^{-1}u)}{k^\pm_s(q^{2(1-s)}\xi^{-1}u)}\,,
\end{equation}
where $1\leq\ell\leq n+1$ for odd $N=2n+1$  and 
$1\leq\ell\leq n$ for $N=2n$ even.

Proposition~\ref{inverse} has an obvious 
\begin{cor}
There are relations between Gauss coordinates of the fundamental $\LL$-operators 
in the algebra $U_q(\gaf)$ corresponding to the simple roots of the underlying  
algebra $\mathfrak{g}$ 
\begin{equation}\label{GCrel1}
\begin{split}
\FF^\pm_{N+1-i,N-i}(u)&=-\FF^\pm_{i+1,i}(q^{-2i}\xi^{-1}u)\,,\\
\EE^\pm_{N-i,N+1-i}(u)&=-\EE^\pm_{i,i+1}(q^{-2i}\xi^{-1}u)
\end{split}
\end{equation}
for $\forall N$ and $1\leq i\leq n-1$ and 
\begin{equation}\label{GCrel2}
\begin{split}
\FF^\pm_{n+2,n+1}(u)&=-q^{1/2}\ \FF^\pm_{n+1,n}(q^{-2n}\xi^{-1}u)\,,\\
\EE^\pm_{n+1,n+2}(u)&=-q^{-1/2}\ \EE^\pm_{n,n+1}(q^{-2n}\xi^{-1}u)
\end{split}
\end{equation}
for $N=2n+1$. 
 \end{cor} 

This corollary together with equalities  \r{Ksol}    defines 
the algebraically independent sets of the generators in each of the algebras 
$U_q(\gaf)$ of the type $\gaf=B^{(1)}_{n}$, $C^{(1)}_{n}$, $D^{(1)}_{n}$ and 
$A^{(2)}_{N-1}$.

\section{New realization of the algebra $U_q(\gaf)$}
\label{nrg}

New realization of the quantum affine algebras $U_q(\widehat{\mathfrak{g}})$ 
was given in \cite{D88} in terms of the formal series called {\it currents} labeled 
by the simple roots of the underlying finite-dimensional algebra $\mathfrak{g}$. 
Relations between currents and Gauss coordinates for the algebra 
$\tilde U_q(A^{(1)}_{N-1})$ was given in \cite{DF93}
\begin{equation}\label{glNcurr} 
\begin{split}
F_i(u)&=\FF^+_{i+1,i}(u)-\FF^-_{i+1,i}(u)=\sum_{\ell\in\ZZ}{{\rm sign}(\ell)}\FF_{i+1,i}[\ell]u^{-\ell} ,\\
E_i(u)&=\EE^+_{i,i+1}(u)-\EE^-_{i,i+1}(u)=-\sum_{\ell\in\ZZ}{{\rm sign}(-\ell)}\EE_{i,i+1}[\ell]u^{-\ell},
\end{split}
\end{equation}
where $1\leq i\leq N-1$. 
 
For the algebras  $U_q(B^{(1)}_{n})$, $U_q(C^{(1)}_{n})$ and 
$U_q(A^{(2)}_{N-1})$ the currents are introduced by the formulas \r{glNcurr} 
for $1\leq i\leq n$. For the algebra $U_q(D^{(1)}_{n})$ first $(n-1)$ currents 
are also introduced by \r{glNcurr} with $1\leq i\leq n-1$ and the currents 
$F_n(u)$ and $E_n(u)$ by the equalities \cite{JLM19a}
\begin{equation}\label{Dcurr}
\begin{split}
F_n(u)&=\FF^+_{n+1,n-1}(u)-\FF^-_{n+1,n-1}(u)=\FF^-_{n+2,n}(u)-\FF^+_{n+2,n}(u)\,,\\
E_n(u)&=\EE^+_{n-1,n+1}(u)-\EE^-_{n-1,n+1}(u)=\EE^-_{n,n+2}(u)-\EE^+_{n,n+2}(u)\,.
\end{split}
\end{equation}

It is obvious from the commutation relations in the algebra $U_q(\gaf)$ \r{TM-1}
that term $\QR(u,v)$ of the $\RR$-matrix \r{R-mat} do not 
contribute into commutation relations of the matrix entries $\LL^\pm_{i,j}(u)$ and 
$\LL^\pm_{k,l}(v)$ for $1\leq i,j,k,l\leq n$. Commutation relations 
between these matrix entries and between corresponding Gauss coordinates 
are defined by $U_q(\mathfrak{gl}_{n})$-invariant $\RR$-matrix \r{R-matA}.
These commutation relations can be translated into commutation relations 
between currents $F_i(u)$, $E_i(u)$, $1\leq i\leq n-1$ and Gauss coordinates 
$k^\pm_{\ell}(u)$, $1\leq\ell\leq n$ according to the standard approach 
developed  in \cite{DF93}. We formulate  all nontrivial commutation relations between these currents without proofs
\begin{equation*}\label{kiFA}
\begin{split}
k^{\pm}_i(u) F_i(v) k^{\pm}_i(u)^{-1}&= \frac{q^{-1}u -qv}{u-v}\   F_i(v),\\
k^{\pm}_{i+1}(u)F_i(v)k^{\pm}_{i+1}(u)^{-1}&= \frac{qu -q^{-1}v}{u-v}\   F_i(v),
\end{split}
\end{equation*}
\begin{equation}\label{kEFA}
\begin{split}
k^{\pm}_i(u)^{-1}E_i(v)k^{\pm}_i(u)&=\frac{q^{-1}u -qv}{u-v}\   E_i(v),\\
k^{\pm}_{i+1}(u)^{-1}E_i(v)k^{\pm}_{i+1}(u)&=\frac{qu -q^{-1}v}{u-v}\  E_i(v),
\end{split}
\end{equation}
\begin{equation*}\label{FiFiA}
(q^{-1}u -qv)\ F_i(u)F_i(v)=  (qu -q^{-1}v)\  F_i(v)F_i(u),
\end{equation*}
\begin{equation*}\label{EiEiA}
(qu -q^{-1}v)\ E_i(u) E_i(v)=  (q^{-1}u -qv)\  E_i(v) E_i(u),
\end{equation*}
\begin{equation*}\label{FiFiiA}
(u-v)\ F_i(u)F_{i+1}(v)= (q^{-1}u-qv)\ F_{i+1}(v)F_i(u),
\end{equation*}
\begin{equation*}\label{EiEiiA}
(q^{-1}u-qv)\ E_i(u)E_{i+1}(v)= (u-v)\  E_{i+1}(v)E_i(u),
\end{equation*}
\begin{equation*}\label{EFA}
[E_i(u),F_j(v)]=\delta_{i,j}\ (q-q^{-1}) 
\delta(u,v)\Big(k^-_{i+1}(v)\,k^-_{i}(v)^{-1}-k^+_{i+1}(u)\,k^{+}_{i}(u)^{-1}\Big).
\end{equation*}
 There are also  Serre relations for the currents $E_i(u)$ and $F_i(u)$ \cite{D88,DF93} 
\begin{equation}\label{Serre}
\begin{split}
\mathop{\rm Sym}_{v_1,v_2}\Big[F_i(v_1),[F_i(v_2),F_{i\pm1}(u)]_{q^{-1}}\Big]_q&=0\,,\\
\mathop{\rm Sym}_{v_1,v_2}\Big[E_i(v_1),[E_i(v_2),E_{i\pm1}(u)]_{q}\Big]_{q^{-1}}&=0\,,
\end{split}
\end{equation}
where  $[A,B]_q$ means $q$-commutator 
\begin{equation*}
[A,B]_q=A\,B-q\ B\,A
\end{equation*}
and $\mathop{\rm Sym}_{v_1,v_2}G(v_1,v_2)\equiv G(v_1,v_2)+G(v_2,v_1)$.

The multiplicative delta function in  \r{kEFA} is defined by the formal series 
\begin{equation*}\label{delta}
\delta(u,v)=\sum_{\ell\in\ZZ} \frac{u^\ell}{v^\ell}
\end{equation*}
which satisfy the property
\begin{equation*}\label{delta-prop}
\delta(u,v)G(u)=\delta(u,v)G(v)
\end{equation*}
for any formal series $G(u)$.

\begin{remark}\label{for-ser}
The equalities \r{kEFA} should be understood in a sense of equalities 
between formal series. It means that these commutation relations should be understood
as infinite set of equalities between modes of the currents which appear after equating 
the coefficients at all powers $u^\ell v^{\ell'}$ for $\ell,\ell'\in\ZZ$. 
The rational functions in the commutation relations  \r{kEFA}
should be understood as series over powers of $v/u$ in the relations containing 
the current $k^+_j(u)$ and over powers of $u/v$ in the relations with 
the current $k^-_j(u)$.
\end{remark}

The commutation relations of the currents $F_n(u)$, $E_n(u)$ and diagonal 
Gauss coordinates will be specific for each of the algebras 
$U_q(B^{(1)}_{n})$, $U_q(C^{(1)}_{n})$, $U_q(D^{(1)}_{n})$ and
$U_q(A^{(2)}_{N-1})$. According to the theorem~\ref{thm-emb}
these commutation relations can be obtained by considering 
algebras of the small rank presented in the appendix~\ref{smrk}. 

\subsection{New realization of the algebra $U_q(B^{(1)}_n)$}
\label{nrBn}

The full set of the  nontrivial commutation relation for the algebra $U_q(B^{(1)}_n)$
is given by the relations \r{kEFA} and \cite{JLM19a}
\begin{equation*}\label{kiFB}
\begin{split}
k^{\pm}_n(u) F_n(v) k^{\pm}_n(u)^{-1}&= \frac{q^{-1}u -qv}{u-v}\ F_n(v),\\
k^{\pm}_{n+1}(u)F_{n}(v)k^{\pm}_{n+1}(u)^{-1}&=\frac{q^{-1}u -qv}{u-v}\ 
\frac{qu-v}{u-qv}\ F_{n}(v),
\end{split}
\end{equation*}
\begin{equation*}\label{kiEB}
\begin{split}
k^{\pm}_n(u)^{-1}E_n(v)k^{\pm}_n(u)&= \frac{q^{-1}u -qv}{u-v}\ E_n(v),\\
k^{\pm}_{n+1}(u)^{-1}E_{n}(v)k^{\pm}_{n+1}(u)&=\frac{q^{-1}u -qv}{u-v}\ 
\frac{qu-v}{u-qv}\ E_{n}(v),
\end{split}
\end{equation*}
\begin{equation*}\label{FiFiB}
(u-qv)\ F_{n}(u)F_{n}(v)=  (qu-v)\  F_{n}(v)F_{n}(u),
\end{equation*}
\begin{equation*}\label{EiEiB}
(qu-v)\ E_{n}(u)E_{n}(v)=  (u-qv)\  E_{n}(v)E_{n}(u),
\end{equation*}
\begin{equation*}\label{FiFiiB}
(u-v)\ F_{n-1}(u)F_{n}(v)= (q^{-1}u-qv)\ F_{n}(v)F_{n-1}(u),
\end{equation*}
\begin{equation*}\label{EiEiiB}
(q^{-1}u-qv)\ E_{n-1}(u)E_{n}(v)= (u-v)\  E_{n}(v)E_{n-1}(u),
\end{equation*}
\begin{equation*}\label{EFB}
[E_n(u),F_n(v)]= (q-q^{-1}) 
\delta(u,v)\Big(k^-_{n+1}(v)\,k^-_{n}(v)^{-1}-k^+_{n+1}(u)\,k^{+}_{n}(u)^{-1}\Big),
\end{equation*}
 where modes of the dependent currents $k^\pm_{n+1}(u)$ are defined by the relation 
\begin{equation*}
\prod_{\ell=1}^n k^\pm_\ell(q^{2(n-\ell)}u)=k^\pm_{n+1}(qu)\ k^\pm_{n+1}(u)\ 
\prod_{\ell=1}^n k^\pm_\ell(q^{2(n-\ell+1)}u)
\end{equation*}
following from \r{Kinv} and \r{Kinvall} for $\ell=n+1$ and $\xi=q^{1-2n}$. 
Serre relations which include currents $E_n(u)$ and $F_n(u)$  are
\begin{equation}\label{SerreB}
\begin{split}
\mathop{\rm Sym}_{v_1,v_2} \Big[F_{n-1}(v_1),[F_{n-1}(v_2),F_{n}(u)]_{q^{-1}}\Big]_q &=0\,,\\
\mathop{\rm Sym}_{v_1,v_2}
\Big[E_{n-1}(v_1),[E_{n-1}(v_2),E_{n}(u)]_{q}\Big]_{q^{-1}} &=0\,,\\
\mathop{\rm Sym}_{v_1,v_2,v_3}\Big[F_n(v_1),\Big[F_n(v_2),[F_n(v_3),F_{n-1}(u)]_{q^{-1}}\Big]_q\Big]&=0\,,\\
\mathop{\rm Sym}_{v_1,v_2,v_3}\Big[E_n(v_1),\Big[E_n(v_2),[E_n(v_3),E_{n-1}(u)]_{q}\Big]_{q^{-1}}\Big]&=0\,.
\end{split}
\end{equation}
Here $\mathop{\rm Sym}_{v_1,v_2,v_3}G(v_1,v_2,v_2)\equiv 
\sum_{\sigma\in \mathfrak{S}_3}G(v_{\sigma(1)},v_{\sigma(2)},v_{\sigma(3)})$.

\subsection{New realization of the algebra  $U_q(C^{(1)}_n)$}
\label{nrCn}

The nontrivial commutation relations for the set of the currents 
in the algebra $U_q(C^{(1)}_n)$ are given by the commutation relations 
\r{kEFA} and the commutation relations involving the 
currents $F_n(u)$ and $E_n(u)$ \cite{JLM19}
\begin{equation*}\label{kiFC}
k^{\pm}_{n}(u)F_{n}(v)k^{\pm}_{n}(u)^{-1}=\frac{q^{-2}u-q^2v}{u-v}\   F_{n}(v),
\end{equation*}
\begin{equation*}\label{kiEC}
k^{\pm}_{n}(u)^{-1}E_{n}(v)k^{\pm}_{n}(u)=\frac{q^{-2}u-q^2v}{u-v}\ E_{n}(v),
\end{equation*}
\begin{equation*}\label{FiFiC}
(q^{-2}u-q^2v)\ F_{n}(u)F_{n}(v)=  (q^{2}u-q^{-2}v)\  F_{n}(v)F_{n}(u),
\end{equation*}
\begin{equation*}\label{EiEiC}
 (q^{2}u-q^{-2}v)\  E_{n}(u)E_{n}(v)=  (q^{-2}u-q^2v)\  E_{n}(v)E_{n}(u),
\end{equation*}
\begin{equation*}\label{tFiFiiC}
(u-v)\ F_{n-1}(u)F_{n}(v)= (q^{-2}u-q^2v)\ F_{n}(v)F_{n-1}(u)
\end{equation*}
\begin{equation*}\label{EiEiiC}
(q^{-2}u-q^2v)\ E_{n-1}(u)E_{n}(v)= (u-v)\  E_{n}(v)E_{n-1}(u),
\end{equation*}
\begin{equation*}\label{tEFC}
[E_n(u),F_n(v)]= (q^{2}-q^{-2}) \delta(u,v)
 \Big(k^-_{n+1}(u)\cdot k^-_{n}(u)^{-1}-k^+_{n+1}(v)\cdot k^+_{n}(v)^{-1}\Big),
\end{equation*}
where 
\begin{equation*}
k^\pm_{n+1}(u)=k^\pm_{n}(q^{4}u)^{-1}\prod_{\ell=1}^{n-1}
\frac{k^\pm_{\ell}(q^{2n+2-2\ell}u)}{k^\pm_{\ell}(q^{2n+4-2\ell}u)}\,.
\end{equation*}
Serre relations which includes the currents $F_n(u)$ and $E_n(u)$   are  
\begin{equation}\label{SerreC}
\begin{split}
\mathop{\rm Sym}_{v_1,v_2} \Big[F_{n}(v_1),[F_{n}(v_2),F_{n-1}(u)]_{q^{-2}}\Big]_{q^2} &=0\,,\\
\mathop{\rm Sym}_{v_1,v_2}
\Big[E_{n}(v_1),[E_{n}(v_2),E_{n-1}(u)]_{q^{2}}\Big]_{q^{-2}} &=0\,,\\
\mathop{\rm Sym}_{v_1,v_2,v_3}\Big[F_{n-1}(v_1),\Big[F_{n-1}(v_2),[F_{n-1}(v_3),F_{n}(u)]_{q^{-2}}\Big]_{q^2}\Big]&=0\,,\\
\mathop{\rm Sym}_{v_1,v_2,v_3}\Big[E_{n-1}(v_1),\Big[E_{n-1}(v_2),[E_{n-1}(v_3),E_{n}(u)]_{q^{2}}\Big]_{q^{-2}}\Big]&=0\,.
\end{split}
\end{equation}

\subsection{New realization of the algebra  $U_q(D^{(1)}_n)$}
\label{nrDn}

Using results presented in appendix~\ref{o2n} one can obtain that the 
current realization of the algebra $U_q(D^{(1)}_n)$ is given by the 
commutation relations \r{kEFA} and all nontrivial commutation  relations 
which includes the currents $F_n(u)$ and $E_n(u)$ are \cite{JLM19a}
\begin{equation*}\label{kiFD}
\begin{split}
k^{\pm}_{n-1}(u)F_{n}(v)k^{\pm}_{n-1}(u)^{-1}&=\frac{q^{-1}u -qv}{u-v}\  F_{n}(v),\\
k^{\pm}_{n}(u)F_{n}(v)k^{\pm}_{n}(u)^{-1}&=\frac{q^{-1}u -qv}{u-v}\  F_{n}(v),
\end{split}
\end{equation*}
\begin{equation*}\label{tkiED}
\begin{split}
k^{\pm}_{n-1}(u)^{-1}E_{n}(v)k^{\pm}_{n-1}(u)&=\frac{q^{-1}u -qv}{u-v}\ E_{n}(v),\\
k^{\pm}_{n}(u)^{-1}E_{n}(v)k^{\pm}_{n}(u)&=\frac{q^{-1}u -qv}{u-v}\ E_{n}(v),
\end{split}
\end{equation*}
\begin{equation*}\label{tFiFiD}
(q^{-1}u-qv)\ F_n(u)F_n(v)=  (qu-q^{-1}v)\  F_n(v)F_n(u),
\end{equation*}
\begin{equation*}\label{tEiEiD}
(qu-q^{-1}v)\ E_n(u) E_n(v)=  (q^{-1}u-qv)\  E_n(v) E_n(u),
\end{equation*}
\begin{equation*}\label{tFiFiiD}
(u-v)\ F_{n-2}(u)F_{n}(v)=  (q^{-1}u-qv)\ F_{n}(v)F_{n-2}(u),
\end{equation*}
\begin{equation*}\label{tEiEiiD}
(q^{-1}u-qv)\ E_{n-2}(u)E_{n}(v)= (u-v)\  E_{n}(v)E_{n-2}(u),
\end{equation*}
\begin{equation*}\label{tEFD}
[E_n(u),F_n(v)]= (q-q^{-1})\delta(u,v)\Big(k^-_{n+1}(u)\cdot k^-_{n-1}(u)^{-1}-
k^+_{n+1}(v)\cdot k^+_{n-1}(v)^{-1}\Big).
\end{equation*}
In \r{tEFD} the Gauss coordinates $k^\pm_{n+1}(u)$ are given by \r{Ksol} for 
$\ell=n$ and $\xi=q^{2-2n}$.
 The Serre relations which includes currents
$F_n(u)$ and $E_n(u)$  can be written in the form  \cite{JLM19a}
\begin{equation*}\label{SerreD}
\begin{split}
\mathop{\rm Sym}_{v_1,v_2}\Big[F_{i}(v_1),[F_{i}(v_2),F_{n}(u)]_{q^{-1}}\Big]_q&=0,
\quad \mathop{\rm Sym}_{v_1,v_2}\Big[F_{n}(v_1),[F_{n}(v_2),F_{i}(u)]_{q^{-1}}\Big]_q=0\,,\\
\mathop{\rm Sym}_{v_1,v_2}\Big[E_{i}(v_1),[E_{i}(v_2), E_{n}(u)]_{q}\Big]_{q^{-1}}&=0,
\quad \mathop{\rm Sym}_{v_1,v_2}\Big[E_{n}(v_1),[E_{n}(v_2),E_{i}(u)]_{q}\Big]_{q^{-1}}=0,
\end{split}
\end{equation*}
for $i=n-2,n-1$.

\subsection{New realization of the algebra  $U_q(A^{(2)}_{2n})$}
\label{nrA2n}

 Nontrivial commutation relations for the new realization of the algebra $U_q(A^{(2)}_{2n})$ is given by 
 the relations \r{kEFA} and additional  relations 
which includes  currents $F_n(u)$ and $E_n(u)$ listed below (most of them was found in case $U_q(A^{(2)}_{2})$ \cite{Sh10})
\begin{equation*}\label{kiFA2n}
\begin{split}
k^{\pm}_n(u) F_n(v) k^{\pm}_n(u)^{-1}&= \frac{q^{-1}u -qv}{u-v}\ F_n(v),\\
k^{\pm}_{n+1}(u)F_{n}(v)k^{\pm}_{n+1}(u)^{-1}&=\frac{qu -q^{-1}v}{u-v}\ 
\frac{qv+u}{v+qu}\ F_{n}(v),
\end{split}
\end{equation*}
\begin{equation*}\label{kiEA2n}
\begin{split}
k^{\pm}_n(u)^{-1}E_n(v)k^{\pm}_n(u)&=\frac{q^{-1}u -qv}{u-v}\ E_n(v),\\
k^{\pm}_{n+1}(u)^{-1}E_{n}(v)k^{\pm}_{n+1}(u)&=\frac{qu -q^{-1}v}{u-v}\ \frac{qv+u}{v+qu}\ E_{n}(v),
\end{split}
\end{equation*}
\begin{equation*}\label{FiFiA2n}
(q^{-1}u-qv)(qu+v)\ F_{n}(u)F_{n}(v)=  (qu-q^{-1}v)(u+qv)\  F_{n}(v)F_{n}(u),
\end{equation*}
\begin{equation*}\label{EiEiA2n}
(qu-q^{-1}v)(u+qv)\ E_{n}(u)E_{n}(v)=  (q^{-1}u-qv)(qu+v)\  E_{n}(v)E_{n}(u),
\end{equation*}
\begin{equation*}\label{FiFiiA2n}
(u-v)\ F_{n-1}(u)F_{n}(v)= (q^{-1}u-qv)\ F_{n}(v)F_{n-1}(u),
\end{equation*}
\begin{equation*}\label{EiEiiA2n}
(q^{-1}u-qv)\ E_{n-1}(u)E_{n}(v)= (u-v)\  E_{n}(v)E_{n-1}(u),
\end{equation*}
\begin{equation*}\label{EFA2n}
[E_n(u),F_n(v)]= (q-q^{-1}) 
\delta(u,v)\Big(k^-_{n+1}(v)\,k^-_{n}(v)^{-1}-k^+_{n+1}(u)\,k^{+}_{n}(u)^{-1}\Big),
\end{equation*}
where $k^\pm_{n+1}(u)$ are defined by the relation 
\begin{equation*}
k^\pm_{n+1}(u)k^\pm_{n+1}(-qu)=\prod_{\ell=1}^{n}
\frac{k^\pm_\ell(-q^{2n-2\ell+1}u)}{k^\pm_\ell(-q^{2n-2\ell+3}u)}
\end{equation*}
following from \r{Ksol} at $\ell=n+1$ and $\xi=-q^{-1-2n}$. 
Serre relations which include currents   $E_n(u)$ and $F_n(u)$ are the same as in the case of $U_q(B^{(1)}_n)$ \r{SerreB} 
\begin{equation*}\label{SerreA2even1}
\begin{split}
\mathop{\rm Sym}_{v_1,v_2} \Big[F_{n-1}(v_1),[F_{n-1}(v_2),F_{n}(u)]_{q^{-1}}\Big]_q &=0\,,\\
\mathop{\rm Sym}_{v_1,v_2}
\Big[E_{n-1}(v_1),[E_{n-1}(v_2),E_{n}(u)]_{q}\Big]_{q^{-1}} &=0\,,\\
\mathop{\rm Sym}_{v_1,v_2,v_3}\Big[F_n(v_1),\Big[F_n(v_2),[F_n(v_3),F_{n-1}(u)]_{q^{-1}}\Big]_q\Big]&=0\,,\\
\mathop{\rm Sym}_{v_1,v_2,v_3}\Big[E_n(v_1),\Big[E_n(v_2),[E_n(v_3),E_{n-1}(u)]_{q}\Big]_{q^{-1}}\Big]&=0\,.
\end{split}
\end{equation*}
and there are additional Serre relations 
for the currents $E_n(u)$ and $F_n(u)$ which can be presented in the form \cite{D88, Sh10}
\begin{equation*}\label{SerreA2even2}
\begin{split}
\mathop{\rm Sym}_{u,v,w}\big(u-(q+q^2)v+q^3w\big)F_n(u)F_n(v)F_n(w)&=0\,,\\
\mathop{\rm Sym}_{u,v,w}\big(q^3vw-(q+q^2)uw+uv\big)F_n(u)F_n(v)F_n(w)&=0\,,\\
\mathop{\rm Sym}_{u,v,w}\big(q^3u-(q+q^2)v+w\big)E_n(u)E_n(v)E_n(w)&=0\,,\\
\mathop{\rm Sym}_{u,v,w}\big(vw-(q+q^2)uw+q^3uv\big)E_n(u)E_n(v)E_n(w)&=0\,.
\end{split}
\end{equation*}

\subsection{New realization of the algebra  $U_q(A^{(2)}_{2n-1})$}
\label{nrA2n1}

Finally using results presented 
in appendix~\ref{A2n1} one can describe 
the new realization of the algebra $U_q(A^{(2)}_{2n-1})$ as collection of the commutation 
relations \r{kEFA} and additional  relations 
which includes currents $F_n(u)$ and $E_n(u)$
\begin{equation*}\label{A2n-1-2}
k^{\pm}_{n}(u)F_{n}(v)k^{\pm}_{n}(u)^{-1}=\frac{q^2v^2-q^{-2}u^2}{v^2-u^2}F_{n}(v),
\end{equation*}
\begin{equation*}\label{A2n-1-3}
k^{\pm}_{n}(u)^{-1}E_{n}(v)k^{\pm}_{n}(u)=\frac{q^2v^2-q^{-2}u^2}{v^2-u^2}E_{n}(v),
\end{equation*}
\begin{equation*}\label{A2n-1-4}
(q^{-2}u^2-q^2v^2)\ F_{n}(u)F_{n}(v)= (q^{2}u^2-q^{-2}v^2) \  F_{n}(v)F_{n}(u),
\end{equation*}
\begin{equation*}\label{A2n-1-5}
(q^{2}u^2-q^{-2}v^2)\ E_{n}(u)E_{n}(v)=  (q^{-2}u^2-q^2v^2)\  E_{n}(v)E_{n}(u),
\end{equation*}
\begin{equation*}\label{A2n-1-6}
(u^2-v^2)\ F_{n-1}(u)F_{n}(v)= (q^{-2}u^2-q^2v^2)\ F_{n}(v)F_{n-1}(u),
\end{equation*}
\begin{equation*}\label{A2n-1-7}
(q^{-2}u^2-q^2v^2)\ E_{n-1}(u)E_{n}(v)= (u^2-v^2)\  E_{n}(v)E_{n-1}(u),
\end{equation*}
\begin{equation*}\label{A2n-1-9}
[E_{n}(u),F_{n}(v)]=(q^2-q^{-2})\bar\delta(u,v)\sk{k^-_{n+1}(v)k^-_n(v)^{-1}-
k^+_{n+1}(u)k^+_n(u)^{-1}},
\end{equation*}
where $\delta$-function $\bar\delta(u,v)$ is given by the series
\begin{equation*}
\bar\delta(u,v)=\sum_{\ell\in\ZZ}\frac{u^{2\ell +1}}{v^{2\ell +1}}\,.
\end{equation*}
The currents $F_n(u)$ and $E_n(u)$ are series with respect to the odd 
powers of the spectral parameters
and the ratio of the Gauss  coordinates $k^\pm_{n+1}(u)k^\pm_{n}(u)^{-1}$
are series with respect to even powers of the spectral parameters. 
 Serre relations which includes the currents $F_n(u)$ and $E_n(u)$   are  
 the same as in the case of $U_q(C^{(1)}_n)$ \r{SerreC}.
\begin{equation*}\label{SerreA2odd}
\begin{split}
\mathop{\rm Sym}_{v_1,v_2} \Big[F_{n}(v_1),[F_{n}(v_2),F_{n-1}(u)]_{q^{-2}}\Big]_{q^2} &=0\,,\\
\mathop{\rm Sym}_{v_1,v_2}
\Big[E_{n}(v_1),[E_{n}(v_2),E_{n-1}(u)]_{q^{2}}\Big]_{q^{-2}} &=0\,,\\
\mathop{\rm Sym}_{v_1,v_2,v_3}\Big[F_{n-1}(v_1),\Big[F_{n-1}(v_2),[F_{n-1}(v_3),F_{n}(u)]_{q^{-2}}\Big]_{q^2}\Big]&=0\,,\\
\mathop{\rm Sym}_{v_1,v_2,v_3}\Big[E_{n-1}(v_1),\Big[E_{n-1}(v_2),[E_{n-1}(v_3),E_{n}(u)]_{q^{2}}\Big]_{q^{-2}}\Big]&=0\,.
\end{split}
\end{equation*}

\section{Currents and  the projections}
\label{curproj}

Currents and  the Gauss coordinates can be related through projections $\Pfpm$ 
and $\Pepm$ onto subalgebras $U^\pm_f$ and $U^\pm_e$
acting on the   ordered products  of the currents. Rigorous definitions 
of these projections depends on the type of the cycling ordering 
 of the Cartan-Weyl generators  (see \cite{EKhP07} for 
detailed exposition of the properties of the projections for the ordering \r{order1}). 

Denote by  $\overline U_f$  \cite{DKhP00-2}
an extension of the algebra $U_f=U^-_f\cup U^+_f\cup U^+_k$ formed
by linear combinations of series, given as infinite sums of monomials
$a_{i_1}[n_1]\cdots a_{i_k}[n_k]$ with $n_1\leq\cdots\leq n_k$, and $n_1+...+n_k$
fixed, where  $a_{i_l}[n_l]$ is either $F_{i_l}[n_l]$ or $k^+_{i_l}[n_l]$.
Analogously, denote by  
$\overline U_e$ an extension of the algebra $U_e=U^-_e\cup U^+_e\cup U^-_k$ formed
by linear combinations of series, given as infinite sums of monomials
$a_{i_1}[n_1]\cdots a_{i_k}[n_k]$ with $n_1\geq\cdots\geq n_k$, and $n_1+...+n_k$
fixed, where  $a_{i_l}[n_l]$ is either $E_{i_l}[n_l]$ or $k^-_{i_l}[n_l]$.

It was proved in  \cite{EKhP07,KhP-Kyoto,DKhP00-2} that the ordered products 
of the simple roots currents $F_{j-1}(u)\cdots F_i(u)$ and 
$E_{i}(u)\cdots E_{j-1}(u)$ in the algebra $\tilde U_q(A^{(1)}_{N-1})$ 
are well defined and belong to $\overline U_f$ and 
$\overline U_e$ respectively. Moreover, the actions of the projections 
$\Pfpm$ and $\Pepm$ onto these elements are well defined. 

The action of the projections $\Pfpm$ onto product of the 
currents  can be defined as follows.  
In order to calculate projections from such product 
one has to substitute each current by the difference 
of the corresponding Gauss coordinates \r{glNcurr} and using the commutation 
relations between them order the product of the currents $F_i(u)$ in a way that 
all negative Gauss coordinates $\FF^-_{j,i}(u)$ will be on the left of all positive 
Gauss coordinates $\FF^+_{k,l}(v)$. Then application of the projection
$\Pfp$ to this product of the currents is removing all the terms which have 
at least one negative Gauss coordinate on the left. Analogously, application 
of the projection $\Pfm$ is removing all the terms which have at least 
one positive Gauss coordinate on the right. The action of the projections 
$\Pepm$ onto product of the currents $E_i(u)$ is defined analogously, 
but the ordering of the Gauss coordinates is inverse: all positive 
Gauss coordinates $\EE^+_{i,j}(u)$ should be placed on the left of 
all negative coordinates $\EE^-_{l,k}(v)$ using the commutation relations between 
them and according to the ordering \r{order1}.

For $U_q(B^{(1)}_{n})$, $U_q(C^{(1)}_{n})$, $U_q(D^{(1)}_{n})$ and 
$U_q(A^{(2)}_{N-1})$ we introduced currents $F_i(u)$ and $E_i(u)$,
 $1\leq i\leq n$  by the formulas \r{glNcurr} and \r{Dcurr}. Using relations \r{GCrel1}
we define dependent currents $F_i(u)$ and $E_i(u)$,
$n'\leq i\leq N-1$ 
\begin{equation}\label{depcur}
F_i(u)=-F_{N-i}(q^{2(i-N)}\xi^{-1}u)\,,\quad 
E_i(u)=-E_{N-i}(q^{2(i-N)}\xi^{-1}u)\,.
\end{equation}

For the algebras $U_q(B^{(1)}_{n})$ and $U_q(A^{(2)}_{2n})$
and according to \r{GCrel2} we introduce  additional  dependent  currents  
\begin{equation}\label{depcur2}
F_{n+1}(u)=-q^{1/2}\  F_{n}(q^{-2n}\xi^{-1}u)\,,\quad 
E_{n+1}(u)=-q^{-1/2}\ E_{n}(q^{-2n}\xi^{-1}u)\,.
\end{equation}

For $1\leq i<j\leq N$ one can define 
the elements $\eF_{j,i}(u)$ and $\eE_{i,j}(u)$ from the completed subalgebras $\overline U_f$
and $\overline U_e$ 
\begin{equation}\label{CoCu1}
\begin{split}
\eF_{j,i}(u)&= 
F_{j-1}(u)F_{j-2}(u)\cdots F_{i+1}(u)F_i(u),\\
\eE_{i,j}(u)&= 
E_{i}(u)E_{i+1}(u)\cdots E_{j-2}(u)E_{j-1}(u) ,
\end{split}
\end{equation}
for $U_q(B^{(1)}_n)$, $U_q(C^{(1)}_n)$, $U_q(A^{(2)}_{N-1})$ and
\begin{equation*}\label{CoCu1F}
\eF_{j,i}(u)=\begin{cases}
&F_{j-1}(u)F_{j-2}(u)\cdots F_{i+1}(u)F_i(u), \quad i<j\leq n \quad\text{or}\quad n+1\leq i<j,\\
&0, \quad \quad \quad \quad \quad \quad \quad\quad \quad \quad \quad \quad \quad \quad   i=n,\; j=n+1,\\
&F_n(u)F_{n-2}(u)\cdots F_i(u), \quad \quad \quad \quad \quad i\leq n-1,\; j=n+1,\\
&-F_{j-1}(u)\cdots F_{n+2}(u) F_n(u), \quad \quad\quad\; i=n,\; j\geq n+2,\\
&F_{j-1}(u)\cdots F_{n+2}(u)F_{n+1}(u)F_n(u)F_{n-2}(u)\cdots F_i(u), \quad {i<n<j-1}
\end{cases}
\end{equation*}
\begin{equation*}\label{CoCu1E}
\eE_{i,j}(u)=\begin{cases}
&E_{i}(u)E_{i+1}(u)\cdots E_{j-2}(u)E_{j-1}(u), \quad i<j\leq n \quad\text{or}\quad n+1\leq i<j,\\
&0, \;\;\quad \quad \quad \quad \quad \quad \quad\quad \quad \quad \quad \quad \quad \quad   i=n,\; j=n+1,\\
&E_{i}(u) \cdots E_{n-2}(u)E_{n}(u), \quad \quad \quad \quad \quad i\leq n-1,\; j=n+1,\\
&-E_{n}(u)E_{n+2}(u)\cdots E_{j-1}(u), \quad \quad\quad\; i=n,\; j\geq n+2,\\
&E_{i}(u)\cdots E_{n-2}(u)E_{n}(u)E_{n+1}(u)E_{n+2}(u)\cdots E_{j-1}(u), \quad 
{i<n<j-1}
\end{cases}
\end{equation*}
for $U_q(D^{(1)}_n)$. To define composed currents for $U_q(D^{(1)}_n)$ one can 
use commutativity
$[F_{n-1}(u),F_n(v)]=0$, $[E_{n-1}(u),E_n(v)]=0$
and 
$F_{n+1}(u)=-F_{n-1}(u)$, $E_{n+1}(u)=-E_{n-1}(u)$  (see \r{depcur}).

We call  elements $\eF_{j,i}(u)$ and $\eE_{i,j}(u)$ {\it the composed currents}. 
For the algebra $U_q(A^{(1)}_{N-1})$  these currents were 
investigated in \cite{EKhP07,KhP-Kyoto}.  It was shown there that the analytical 
properties of the products 
of the composed  currents considered in the category of the highest weight representations
are equivalent to the Serre relations for 
the simple root currents.

Using result of \cite{EKhP07} that action of the projections 
$\Pfpm$ and $\Pepm$ can be prolonged  to the extensions $\overline U_f$ and 
$\overline U_e$ respectively we formulate following 
\begin{prop}\label{cvsg1}
There are relations between Gauss coordinates of the fundamental 
$\LL$-operators and projections of the composed currents in the algebra 
$U_q(\gaf)$
\begin{equation}\label{PrvsGc}
\begin{split}
\Pfp\big(\eF_{j,i}(u)\big)&=\FF^+_{j,i}(u),\qquad
\Pfm\big(\eF_{j,i}(u)\big)=\tFF^-_{j,i}(u),\\
\Pep\big(\eE_{i,j}(u)\big)&=\EE^+_{i,j}(u),\qquad  
\Pem\big(\eE_{i,j}(u)\big)=\tEE^-_{i,j}(u).
\end{split}
\end{equation}
\end{prop}

Proof of this proposition will be given in appendix~\ref{prf71} 
simultaneously for all algebras 
$U_q(\gaf)$ and is based on induction over rank  $n$ of these algebras. 
The base of induction is a verification 
of \r{PrvsGc} for all algebras $U_q(\gaf)$ of small ranks performed 
in the appendices~\ref{o2n1}, \ref{sp2n}, \ref{o2n}, \ref{A22n} and \ref{A2n1}.
In particular,  formulas \r{D14} and  \r{D15} are 
base of induction to prove  proposition~\ref{cvsg1} 
in case of the algebra $U_q(D^{(1)}_n)$. 

\section*{Conclusion}

In this paper we investigate quantum loop algebras for all classical series 
(except $U_q(D^{(2)}_n)$) associated to the quantum $\RR$-matrices 
found in~\cite{Jimbo86}. Results obtained in this paper can be used for 
investigation of the space of states of the quantum integrable models 
with the different symmetries of the high rank. This investigation can be 
performed in the framework of the approach to integrable models
proposed and developed in \cite{EKhP07,KhP-Kyoto,PRS}.
In this method the states of  integrable models are expressed 
through current generators of the quantum loop algebras.
To investigate different physical quantities in such models such as scalar
products of the states and form-factors of the local operators 
it is not necessary to have explicit form of the states in terms of the 
current generators. Usually, it is sufficient to get the action 
of monodromy matrix entries onto these states. 
This approach was called {\it zero modes method} and was already used 
in  \cite{HLPRS-RA,LPRS19a,LP21} to investigate the space of states 
in quantum integrable models related to Yangian doubles and 
rational $\mathfrak{g}$-invariant $\RR$-matrices. Using results of 
the present paper we plan to develop this method  
for the integrable models associated with 
$U_q(\mathfrak{g})$-invariant $\RR$-matrices. 

\section*{Acknowledgments}
The work was performed at the Steklov Mathematical Institute of 
Russian Academy of Sciences, Moscow.
This work is supported by the Russian Science Foundation under grant 19-11-00062.

\appendix

\section{Proofs of the lemmas~\ref{lem51} and \ref{lem52}}
\label{prf-lem}

Recall that we denote by $|i\>$ and $\<j|$, $1\leq i,j\leq N$ 
sets of orthonormal vectors in $\CC^N$ with pairing $\CC^N\ot\CC^N\to\CC$: $\<i|j\>=\delta_{ij}$.
Consider $\RR$-matrix \r{R-mat} for $v=q^2u$: $\RR(1,q^2)=\AR(1,q^2)+\Qer(1,q^2)$, where
\begin{equation*}\label{Em12}
\AR(1,q^2)=\sum_{i\not=j}^N\E_{ii}\ot\E_{jj}-q^{-1}\sum_{i<j}^N \E_{ij}\ot\E_{ji}
-q\sum_{i<j}^N \E_{ji}\ot\E_{ij}
\end{equation*}
and calculate 
\begin{equation}\label{Em13}
\RR(1,q^2)|\ell,\ell\>=0\quad \<\ell,\ell|\RR(1,q^2)=0\quad
 \mbox{for}\quad 1\leq \ell\leq N\quad \mbox{and}\quad 
\ell\not=\ell'\,,
\end{equation}
and 
\begin{equation}\label{Em1414}
\begin{split}
\RR(1,q^2)|1,\ell\>&=|1,\ell\>-q|\ell,1\>,\quad 
\RR(1,q^2)|\ell,1\>=|\ell,1\>-q^{-1}|1,\ell\>,\\
\<\ell,1|\RR(1,q^2)&=\<\ell,1|-q\<1,\ell|,\quad 
\<1,\ell|\RR(1,q^2)=\<1,\ell|-q^{-1}\<\ell,1|
\end{split}
\end{equation}
for $1<\ell<N$. 
Equation \r{Em1414} implies
\begin{equation}\label{Em14}
\RR(1,q^2)\big(|1,\ell\>+q|\ell,1\>\big)=0\quad \mbox{for}\quad 1<\ell<N.
\end{equation}
Here $|i,j\>=|i\>\ot|j\>$ and $\<i,j|=\<i|\ot\<j|$ are vectors from 
$(\CC^N)^{\ot2}$.

Consider commutation relation  \r{RLL} at $v=q^2u$
\begin{equation}\label{RLLred}
\RR_{12}(1,q^2)\ \LL^{(1)}(u)\LL^{(2)}(q^2u)=\LL^{(2)}(q^2u)\LL^{(1)}(u)\ \RR_{12}(1,q^2)\,.
\end{equation}

Using  \r{RLL} one can obtain the 
commutation relations between $\LL$-operators $\LL(u)$ and $\Ll(v)$
\begin{equation}\label{Em9}
\RR_{12}(u,v)\RR_{13}(u,q^2v)\ \LL^{(1)}(u)\ \Ll^{(2,3)}(v)=
 \Ll^{(2,3)}(v)\ \LL^{(1)}(u)\ \RR_{13}(u,q^2v)\RR_{12}(u,v)\,,
\end{equation}
where Yang-Baxter equation \r{YB}
\begin{equation*}\label{Em10}
\RR_{12}(u,v)\cdot\RR_{13}(u,q^2v)\cdot\RR_{23}(1,q^2)=\RR_{23}(1,q^2)\cdot
\RR_{13}(u,q^2v)\cdot\RR_{12}(u,v)
\end{equation*}
is used. 

Analogously, one can obtain the commutation relations between $\LL$-operators 
$\Ll(u)$ and $\Ll(v)$
\begin{equation}\label{Em11}
\begin{split}
&\RR_{23}(q^2u,v)\RR_{13}(u,v)\RR_{24}(u,v)\RR_{14}(u,q^2v)\ \Ll^{(1,2)}(u)\ \Ll^{(3,4)}(v)=\\
&\quad =\Ll^{(3,4)}(v)\ \Ll^{(1,2)}(u)\ \RR_{14}(u,q^2v)\RR_{24}(u,v)
\RR_{13}(u,v)\RR_{23}(q^2u,v)\,.
\end{split}
\end{equation}

To prove  lemma~\ref{lem51} one has to obtain two equalities for $1<\ell<N$ 
\begin{equation}\label{Em15}
\RR_{23}(1,q^2)\RR_{13}(u,q^2v)\RR_{12}(u,v)|1,\ell,1\>=f(u,v)
\RR_{23}(1,q^2)|1,\ell,1\>
\end{equation}
and 
\begin{equation}\label{Em16}
\<1,\ell,1|\RR_{12}(u,v)\RR_{13}(u,q^2v)\RR_{23}(1,q^2)=f(u,v)
\<1,\ell,1|\RR_{23}(1,q^2)\,.
\end{equation}
One can verify equality \r{Em15}
\begin{equation*}
\begin{split}
&\RR_{23}(1,q^2)\RR_{13}(u,q^2v)\RR_{12}(u,v)|1,\ell,1\>=\\
&\quad= \RR_{23}(1,q^2)\RR_{13}(u,q^2v)\Big(|1,\ell,1\>+\pf_{\ell1}(u,v)|\ell,1,1\>)\Big)=\\
&\quad= \RR_{23}(1,q^2)\Big((1+\pf_{11}(u,q^2v))|1,\ell,1\>+
\pf_{1\ell}(u,q^2v)\pf_{\ell1}(u,v)|1,1,\ell\>)\Big)=\\
&\quad=\Big(1+\pf_{11}(u,q^2v) -q\,
\pf_{1\ell}(u,q^2v)\pf_{\ell1}(u,v)\Big)\RR_{23}(1,q^2)|1,\ell,1\>=\\
&\quad =f(u,v)\RR_{23}(1,q^2)|1,\ell,1\>\,,
\end{split}
\end{equation*}
where  identity 
\begin{equation*}
f(u,q^2v)-q\,\gle(u,q^2v)\gri(u,v)=f(u,v)
\end{equation*}
and equation \r{Em14} were used. Equality \r{Em16} can be 
checked analogously.

Multiplying  equality \r{Em9} from the left by the vector $\<1,i,1|$ and 
from the right by the vector $|1,j,1\>$ for $1<i,j<N$ and using \r{Em15} 
and \r{Em16} one obtains
\begin{equation*}\label{Em17}
\<1,i,1|\LL^{(1)}(u)\Ll^{(2,3)}(v) |1,j,1\> = 
\<1,i,1|\Ll^{(2,3)}(v)\LL^{(1)}(u) |1,j,1\>
\end{equation*} 
which implies the statement of the lemma \r{lem51}. \qed

To prove equality \r{Em18} of the lemma~\ref{lem52} one can present its left hand 
side 
\begin{equation*}\label{Em25}
\RR^n_{12}(1,q^2)\RR^n_{34}(1,q^2)\RR^n_{14}(u,q^2v)\RR^n_{13}(u,v)|i,1,j,1\>
\end{equation*}
as sum of two terms using \r{R-mat}: 
$\RR^n_{13}(u,v)=\AR^n_{13}(u,v)+\QR^n_{13}(u,v)$.
First term is equal to
\begin{equation}\label{Em26}
\begin{split}
&\RR^n_{12}(1,q^2)\RR^n_{34}(1,q^2)\RR^n_{14}(u,q^2v)\AR^n_{13}(u,v)|i,1,j,1\>=\\
&\quad= \RR^n_{12}(1,q^2)\RR^n_{34}(1,q^2)\AR^n_{13}(u,v)|i,1,j,1\>=\\
&\quad= \RR^n_{12}(1,q^2)\RR^n_{34}(1,q^2)\AR^{n-1}_{13}(u,v)|i,1,j,1\>
\end{split}
\end{equation}
since $1<i,j<N$. 
Indeed, the action of $\RR^n_{14}(u,q^2v)\AR^n_{13}(u,v)$ onto vector $|i,1,j,1\>$
is 
\begin{equation*}
|i,1,j,1\>+\pf_{ji}(u,v)|j,1,i,1\>+\pf_{1i}(u,q^2v)|1,1,j,i\>+ \pf_{ji}(u,v) \pf_{1j}(u,q^2v)|1,1,i,j\>
\end{equation*}
and last two terms are annihilated by the actions of $\RR_{12}(1,q^2)$ due to \r{Em13}.

Now consider the second term 
\begin{equation*}\label{Em27}
\RR^n_{12}(1,q^2)\RR^n_{34}(1,q^2)\RR^n_{14}(u,q^2v)\QR^n_{13}(u,v)|i,1,j,1\>\,,
\end{equation*}
where by definition \r{QQuv} 
\begin{equation*}\label{Em28}
\QR^n_{13}(u,v)|i,1,j,1\>=\delta_{ij'}\sum_{\ell=1}^N \qf_{\ell j}(u,v)|\ell',1,\ell,1\>\,.
\end{equation*}
Calculating the action 
\begin{equation*}\label{Em29}
\begin{split}
&\RR^n_{14}(u,q^2v)|\ell',1,\ell,1\>=|\ell',1,\ell,1\>+\\
&\quad+ \pf_{1\ell'}(u,q^2v)|1,1,\ell,\ell'\>+\delta_{1\ell}\sum_{m=1}^N
\qf_{m1}(u,q^2v)|m',1,1,m\>
\end{split}
\end{equation*}
one can observe that second term in the right hand side drops out due to the 
action $\RR^n_{12}(1,q^2)$
and  \r{Em13} and by the same reasons the sum 
over $m$ reduces to the sum for $1<m<N$. 
Finally, one gets  
\begin{equation}\label{Em30}
\begin{split}
&\RR^n_{12}(1,q^2)\RR^n_{34}(1,q^2)\RR^n_{14}(u,q^2v)\QR^n_{13}(u,v)|i,1,j,1\>=\\
&= \delta_{ij'}{\RR^n_{12}(1,q^2)\RR^n_{34}(1,q^2) }\sum_{\ell=2}^{N-1}
\Big(\qf_{\ell j}(u,v)|\ell',1,\ell,1\>+\qf_{1j}(u,v)\qf_{\ell1}(u,q^2v)|\ell',1,1,\ell\>\Big)=\\
&= \delta_{ij'}{\RR^n_{12}(1,q^2)\RR^n_{34}(1,q^2) }\sum_{\ell=2}^{N-1}
\Big(\qf_{\ell j}(u,v|\xi)-q\ \qf_{1j}(u,v|\xi)\qf_{\ell1}(u,q^2v|\xi)\Big)|\ell',1,\ell,1\>\,,
\end{split}
\end{equation}
where in the last line of \r{Em30} we used \r{Em14} for $\RR^n_{34}(1,q^2)$ and write
explicitly dependence of the functions $\qf_{ij}(u,v|\xi)$ given by \r{a-fun} on parameter $\xi$.

One can check that for all algebras $\gaf=B^{(1)}_n$, $C^{(1)}_n$, $D^{(1)}_n$ and 
$A^{(2)}_{N-1}$ and corresponding parameters $\xi$ 
given by the table \r{Table} following identity is valid
\begin{equation*}\label{Em31}
\qf_{\ell j}(u,v|\xi)-q\ \qf_{1j}(u,v|\xi)\qf_{\ell1}(u,q^2v|\xi)=\qf_{\ell j}(u,v|q^2\xi)\,.
\end{equation*}
Since multiplication of the parameter $\xi$ by $q^2$ means the change of the rank 
$n\to n-1$ for all algebras $U_q(\gaf)$ (see table \r{Table}) one  concludes
that 
\begin{equation}\label{Em32}
\begin{split}
&\RR^n_{12}(1,q^2)\RR^n_{34}(1,q^2)\RR^n_{14}(u,q^2v)\QR^n_{13}(u,v)|i,1,j,1\>=\\
&\quad=\RR^n_{12}(1,q^2)\RR^n_{34}(1,q^2)\QR^{n-1}_{13}(u,v)|i,1,j,1\>\,.
\end{split}
\end{equation}
Summing \r{Em26} and \r{Em32} we obtain \r{Em18}. Equality \r{Em19} can be 
proved analogously. This concludes the proof of the lemma~\ref{lem52}. \qed

\section{Proof of proposition~\ref{inverse}}
\label{appB}

Equalities
\r{identalt} and \r{GChL} imply that  
\begin{equation}\label{idalt}
\LL_{i,j}(u)=\varepsilon_i \varepsilon_j \ q^{\bar\imath-\bar\jmath}\ 
 \sum_{\ell\leq {\rm min}(i,j)}\tEE_{i',\ell'}(\xi u)\ k_{\ell'}(\xi u)^{-1}\
 \tFF_{\ell',j'}(\xi u)\,.
 \end{equation}
Comparing these expressions for the matrix entries of the fundamental 
$\LL$-operator with \r{Em333} proves
equations \r{Finvall}, \r{Einvall} and \r{Kinvall}
 for $1\leq i<j\leq N$ and $1\leq\ell\leq N$.
Introduce 
 matrix entries $\bMM_{i,j}(u)$ for the 
algebra $U_q^{n-1}(\gaf)$ by the equality 
\begin{equation}\label{Em33}
\LL_{i,j}(u)=\bMM_{i,j}(u)+\bEE_{1,i}(u)\bk_1(u)\bFF_{j,1}(u)
=\bMM_{i,j}(u)+\LL_{i,1}(u)\LL_{1,1}(u)^{-1}\LL_{1,j}(u)\,.
\end{equation}
Entries $\bMM_{i,j}(u)$ for $1< i,j < N$ has Gauss decomposition 
\begin{equation}\label{Em34}
\begin{split}
\bMM_{i,j}(u)&=\sum_{2\leq 
\ell\leq{\rm min}(i,j)} \bEE_{\ell,i}(q^{-2(\ell-1)}u)\ \bk_\ell(q^{-2(\ell-1)}u)\ 
\bFF_{j,\ell}(q^{-2(\ell-1)}u)\\
&=\LL_{i,j}(u)-\LL_{i,1}(u)\LL_{1,1}(u)^{-1}\LL_{1,j}(u)\,.
\end{split}
\end{equation}

Calculating matrix elements of the equality \r{RLLred} 
between vectors $\<1,1|$ and $|1,j\>$
using \r{Em13} and \r{Em1414}
one obtains
\begin{equation}\label{Em39}
\LL_{1,j}(q^2u)\LL_{1,1}(u)=q\ \LL_{1,1}(q^2u)\LL_{1,j}(u)
\end{equation}
and matrix entries \r{Em34} can be written in the form 
\begin{equation}\label{Em3434}
\bMM_{i,j}(u)=\Big(\LL_{i,j}(u)\LL_{1,1}(q^{-2}u)-q\ \LL_{i,1}(u)\LL_{1,j}(q^{-2}u)\Big)
\LL_{1,1}(q^{-2}u)^{-1}\,.
\end{equation}
One can prove commutativity 
\begin{equation*}
\bMM_{i,j}(u)\ \LL_{1,1}(v)= \LL_{1,1}(v)\ \bMM_{i,j}(u),\quad 1<i,j<N
\end{equation*}
in the same way as lemma~\ref{lem51} was proved. 

Multiplying  \r{RLLred} 
from the left and from the right by the 
vectors $\<1,i|$ and $|j,1\>$ for $1<i,j<N$ and  using 
 \r{Em1414} one gets 
\begin{equation*}\label{Em42}
\LL_{i,j}(u)\LL_{1,1}(q^2u)-q\ \LL_{1,j}(u)\LL_{i,1}(q^2u)=
\LL_{i,j}(q^2u)\LL_{1,1}(u)-q\ \LL_{i,1}(q^2u)\LL_{1,j}(u)
\end{equation*}
or  due to  \r{Em8} and \r{Em3434} 
\begin{equation}\label{Em43}
\bMM_{i,j}(q^2u)\LL_{1,1}(q^2u)^{-1}=\MM_{i,j}(u)\LL_{1,1}(u)^{-1}\quad
\mbox{for}\quad 1<i,j<N\,.
\end{equation}

We prove only \r{Finv} and \r{Kinv}. Equality \r{Einv} can be proved analogously. 
Comparing  \r{Gauss1}, \r{Em333} and   \r{idalt} for the 
matrix entry $\LL_{1,1}(u)$ one concludes that 
\begin{equation}\label{kbk}
\bk_1(u)=k_1(u)=k_N(\xi u)^{-1}\,.
\end{equation}
Using $\LL_{1,j}(u)=\FF_{j,1}(u)k_1(u)=\bk_1(u)\bFF_{j,1}(u)$ 
equality \r{Em39} can be rewritten in the form 
\begin{equation}\label{Em40}
\bFF_{j,1}(u)=q \FF_{j,1}(q^{-2}u)=k_1(u)^{-1}\FF_{j,1}(u) k_1(u)\,.
\end{equation}
This yields
\begin{equation}\label{Em41}
\bFF_{j,1}(u)=q\FF_{j,1}(q^{-2}u)=
q^{\bar1-\bar\jmath}
\varepsilon_1\varepsilon_j
\tFF_{N,j'}(\xi u)\quad\mbox{for}\quad 1<j<N
\end{equation}
and \r{Finvall} yields
\begin{equation}\label{Em411}
\bFF_{N,1}(u)=q^{\bar1-\bar N}
\varepsilon_1\varepsilon_N
\tFF_{N,1}(\xi u).
\end{equation}

Consider \r{Em43} for $i=j=2$. It yields
\begin{equation}\label{Em44}
\bk_2(u)=k_2(u)\ k_1(q^2u)k_1(u)^{-1}=k_{N-1}(q^2\xi u)^{-1}\,,
\end{equation}
where the second equality follows from \r{Kinvall}. 
Then consider \r{Em43} for $i=2$ and $2<j<N-1$ to obtain 
\begin{equation*}\label{Em45}
\bMM_{2,j}(u)=\bk_2(q^{-2}u)\bFF_{j,2}(q^{-2}u)=
k_1(u)k_1(q^{-2}u)^{-1}\ \FF_{j,2}(q^{-2}u)k_2(q^{-2}u)
\end{equation*}
which can be presented as 
\begin{equation*}\label{Em46}
\bFF_{j,2}(u)=k_2(u)^{-1}\FF_{j,2}(u)k_2(u)\,.
\end{equation*}

Recall now that according to the theorem~\ref{thm-emb} $\LL$-operator 
$\MM(u)$ satisfy commutation relations \r{Em20} for the algebra  
$U^{n-1}_q(\gaf)$ and we can apply analysis as above to have 
$k_2(u)^{-1}\FF_{j,2}(u)k_2(u)=q\ \FF_{j,2}(q^{-2}u)$ (compare with \r{Em40})
and 
\begin{equation}\label{Em47}
\bFF_{j,2}(u)=q\ \FF_{j,2}(q^{-2}u)=q^{\bar2-\bar\jmath}
\varepsilon_2\varepsilon_j\tFF_{N-1,j'}(q^2\xi u)\quad\mbox{for}\quad 2<j<N-1.
\end{equation}
The second equality in \r{Em47} follows from \r{idalt} as well as 
\begin{equation}\label{Em48}
\bFF_{j,2}(u)=q^{\bar2-\bar\jmath}
\varepsilon_2\varepsilon_{j}\tFF_{N-1,j'}(q^2\xi u),\quad j=N-1,N\,.
\end{equation}
Note that equalities \r{Em47}, \r{Em48} and second equality in \r{Em44}
for the Gauss coordinates of the embedded algebra $U_q^{n-1}(\gaf)$  
repeated the equalities \r{Em41}, \r{Em411} and \r{kbk} respectively with 
the only difference that parameter $\xi$ is replaced by the parameter $q^2\xi$.
According to dependence of $\xi$ on the rank $n$ of the algebra $\gaf$ 
this replacement is equivalent to change of the rank $n\to n-1$. 

Continuing embedding process  and repeating these arguments
for the Gauss coordinates $\bEE_{i,j}(u)$ 
one  proves proposition~\ref{inverse}. \qed

\section{Algebra $U_q(\gaf)$ for small ranks}
\label{smrk}

In this appendix we obtain the commutation relations for the  currents 
$F_n(u)$ and $E_n(u)$ for each of the algebra $U_q(\gaf)$ of the small rank.
Here we will introduce different 
rational functions denoting them by the same notations valid inside of each 
subsection. Hope that this will not lead to misunderstanding.

\subsection{Algebras $U_q(B^{(1)}_1)$ and $U_q(B^{(1)}_2)$}
\label{o2n1}

In order to find commutation relations of the special currents in case 
of the algebra $U_q(B^{(1)}_n)$ we first perform investigation of the 
simplest nontrivial example of the algebra $U_q(B^{(1)}_1)$
as it was done in the paper \cite{LPRS19a}.
In this algebra the algebraically independent series of generators 
are $k^\pm_1(u)$, $\FF^\pm_{2,1}(u)$ and $\EE^\pm_{1,2}(u)$ and 
algebraically dependent generating series are $k^\pm_\ell(u)$, $\ell=2,3$ and 
\begin{equation*}\label{conso3}
\begin{split}
\FF^\pm_{3,2}(u)=-q^{1/2}\FF^\pm_{2,1}(q^{-1}u),\quad 
\EE^\pm_{2,3}(u)=-q^{-1/2}\EE^\pm_{1,2}(q^{-1}u),\\
\FF^\pm_{3,1}(v)=-\frac{\sqrt{q}}{1+q}\FF^\pm_{2,1}(v)^2,\quad 
\EE^\pm_{1,3}(v)=-\frac{\sqrt{q}}{1+q}\EE^\pm_{1,2}(v)^2\,.
\end{split}
\end{equation*}
The modes of  $k^\pm_\ell(u)$, $\ell=2,3$ are defined by the 
relations
\begin{equation*}\label{k-sol}
k_3^\pm(u)=k_1^\pm(qu)^{-1},\quad 
k^\pm_1(u)=k^\pm_2(qu)\ k^\pm_2(u)\ k^\pm_1(q^2u)\,.
\end{equation*}

The commutation relations between Gauss coordinates for the 
algebra $U(B^{(1)}_1)$ are 
\begin{equation}\label{kFEB1}
\begin{split}
k_1(u)\FF_{2,1}(v)k_1(u)^{-1}&= f(v,u)\FF_{2,1}(v)-\gle(v,u)\FF_{2,1}(u)\,,\\
k_1(u)^{-1}\EE_{1,2}(v)k_1(u)&= f(v,u)\EE_{1,2}(v)-\gri(v,u)\EE_{1,2}(u)\,,\\
 [\EE_{1,2}(v), \FF_{2,1}(u)] &= \gle(u,v) \left( k_2(u)k_1(u)^{-1} - k_2(v)k_1(v)^{-1} \right),
\end{split}
\end{equation}
\begin{equation}\label{k2FEB1}
\begin{split}
k_2(u)\FF_{2,1}(v)k_2(u)^{-1}&=f(v,u)f(q^{-1}u,v)\FF_{2,1}(v)+\\
&\qquad+\gle(v,u)\FF_{2,1}(u)+\gri(q^{-1}u,v)\FF_{2,1}(q^{-1}u)\,,\\
k_2(u)^{-1}\EE_{1,2}(v)k_2(u)&=f(v,u)f(q^{-1}u,v)\EE_{1,2}(v)+\\
&\qquad+\gri(v,u)\EE_{1,2}(u)+\gle(q^{-1}u,v)\EE_{1,2}(q^{-1}u)\,,
\end{split}
\end{equation}
\begin{equation}\label{FFEEB1}
\begin{split}
\FF_{2,1}(u)\FF_{2,1}(v)&=f(u,qv)\FF_{2,1}(v)\FF_{2,1}(u)+
\frac{\gle(qv,u)}{1+q}\FF_{2,1}(u)^2+\frac{q\gri(qv,u)}{1+q}\FF_{2,1}(v)^2\,,\\
\EE_{1,2}(u)\EE_{1,2}(v)&=f(v,qu)\EE_{1,2}(v)\EE_{1,2}(u)+
\frac{\gle(qu,v)}{1+q}\EE_{1,2}(u)^2+\frac{q\gri(qu,v)}{1+q}\EE_{1,2}(v)^2\,.
\end{split}
\end{equation}

Restoring upper indices $\pm$ in  \r{FFEEB1} at $u=q^{-1}v$
 \begin{equation*}\label{FFcrp}
\FF^+_{2,1}(q^{-1}v)\FF^\pm_{2,1}(v)=
\frac{1}{1+q^{-1}}\FF^+_{2,1}(q^{-1}v)^2+\frac{1}{1+q}\FF^\pm_{2,1}(v)^2
\end{equation*}
and subtracting one equality from another one gets
\begin{equation}\label{FFccp}
\FF^+_{2,1}(q^{-1}v)F_{1}(v)=
\frac{1}{1+q}\FF^+_{2,1}(v)^2-\frac{1}{1+q}\FF^-_{2,1}(v)^2\,.
\end{equation}
Using \r{FFccp} one can calculate 
 the projection $\Pfp\sk{F_2(u)F_1(u)}$ onto 
subalgebra $U^+_f$ assuming that Gauss coordinates in the product of the currents 
$F_2(u)F_1(u)$ are ordered according to the order \r{order1}. We obtain 
\begin{equation*}\label{GcPc}
\Pfp\sk{F_2(u)F_1(u)}=-\sqrt{q}\ \Pfp\sk{F^+_{2,1}(q^{-1}u)F_1(u)}=
-\frac{\sqrt{q}}{1+q}\FF^+_{2,1}(v)^2=\FF^+_{3,1}(u).
\end{equation*}
This relation together with analogous formulas for the projections 
$\Pfm\sk{F_2(u)F_1(u)}$ and $\Pepm\sk{E_1(u)E_2(u)}$ are base 
of the induction proof of the proposition~\ref{cvsg1} which explains 
the relation between Gauss coordinates and projection of the currents
for the algebra $U_q(B^{(1)}_n)$.

Considering similar commutation relations for the algebra $U_q(B^{(1)}_2)$ 
and using embedding theorem~\ref{thm-emb} one obtains 
besides commutation relations \r{kFEB1}--\r{FFEEB1} for 
the Gauss coordinates $k^\pm_\ell(u)$, $\ell=1,2,3$ with 
$\FF^\pm_{3,2}(u)$ and $\EE^\pm_{2,3}(u)$ also the commutation relations 
of these Gauss coordinates
with $\FF^\pm_{2,1}(u)$ and $\EE^\pm_{1,2}(u)$
\begin{equation*}\label{FFEEB3}
\begin{split}
&\FF_{2,1}(v)\FF_{3,2}(u)=f(u,v)\FF_{3,2}(u)\FF_{2,1}(v)+\\
&\quad+\gri(u,v)\Big(\FF_{3,1}(u)-\FF_{3,2}(u)\FF_{2,1}(u)\Big)-\gle(u,v)\FF_{3,1}(v)\,,\\
&\EE_{2,3}(u)\EE_{1,2}(v)=f(u,v)\EE_{1,2}(v)\EE_{2,3}(u)+\\
&\quad+\gle(u,v)\Big(\EE_{1,3}(u)-\EE_{1,2}(u)\EE_{2,3}(u)\Big)-\gri(u,v)\EE_{1,3}(v)\,.
\end{split}
\end{equation*}

This information is sufficient to obtain for the algebra $U_q(B^{(1)}_n)$
the commutation 
relations of the  currents $F_n(u)$, $E_n(u)$, Gauss coordinates 
$k^\pm_\ell(u)$, $1\leq\ell\leq n+1$ and the currents $F_i(u)$, $E_i(u)$, $1\leq i\leq n-1$
given in section~\ref{nrBn}.

\subsection{Algebra $U_q(C^{(1)}_2)$}
\label{sp2n}

Since algebra $U_q(C^{(1)}_1)$ is not representative we start to consider 
first algebra $U_q(C^{(1)}_2)$. In this algebra the algebraically independent 
generating series are $\FF^\pm_{\ell+1,\ell}(u)$, $\EE^\pm_{\ell,\ell+1}(u)$ and 
$k^\pm_{\ell}$ for $\ell=1,2$. Gauss coordinates 
$\FF^\pm_{2,1}(u)$, $\EE^\pm_{1,2}(u)$, $k^\pm_1(u)$ and $k^\pm_2(u)$ 
form the subalgebra in $U_q(C^{(1)}_2)$  
 isomorphic to the algebra $\tilde U_q(A^{(1)}_1)$ and we 
do not write explicitly commutation relations between them. 

Introduce the rational functions relevant to the considered case
\begin{equation*}\label{fgsp}
\fgo(u,v)=\frac{q^2u-q^{-2}v}{u-v},\quad \gole(u,v)=\frac{(q^2-q^{-2})u}{u-v},\quad
\gori(u,v)=\frac{(q^2-q^{-2})v}{u-v}\,.
\end{equation*} 
The rest commutation relations in $U_q(C^{(1)}_2)$ can be written in the form 
\begin{equation*}\label{CC4F}
k_2(v)\FF_{3,2}(u)k_2(v)^{-1}=\fgo(u,v)\FF_{3,2}(u)-\gole(u,v)\FF_{3,2}(v)\,,
\end{equation*}
\begin{equation*}\label{CC4E}
k_2(v)^{-1}\EE_{2,3}(u)k_2(v)=\fgo(u,v)\EE_{2,3}(u)-\gori(u,v)\EE_{2,3}(v)\,,
\end{equation*}
\begin{equation*}\label{CCFi}
\begin{split}
&\fgo(v,u)\FF_{3,2}(u)\FF_{3,2}(v)=\fgo(u,v) \FF_{3,2}(v)\FF_{3,2}(u)+\\
&\qquad+\gole(v,u)\FF_{3,2}(u)^2- \gole(u,v)\FF_{3,2}(v)^2\,,
\end{split}
\end{equation*}
\begin{equation*}\label{CCEi}
\begin{split}
&\fgo(u,v)\EE_{2,3}(u)\EE_{2,3}(v)=\fgo(v,u) \EE_{2,3}(v)\EE_{2,3}(u)+\\
&\qquad+\gori(u,v)\EE_{2,3}(v)^2- \gori(v,u)\EE_{2,3}(u)^2\,,
\end{split}
\end{equation*}
\begin{equation}\label{CC11}
\begin{split}
&\FF_{2,1}(v)\FF_{3,2}(u)=\fgo(u,v)\FF_{3,2}(u)\FF_{2,1}(v)+\\
&\quad +\gori(u,v)\Big(\FF_{3,1}(u)-\FF_{3,2}(u)\FF_{2,1}(u)\Big)-
\gole(u,v)\FF_{3,1}(v)\,,
\end{split}
\end{equation}
\begin{equation*}\label{CC12}
\begin{split}
&\EE_{2,3}(u)\EE_{1,2}(v)=\fgo(u,v)\EE_{1,2}(v)\EE_{2,3}(u)+\\
&\quad +\gole(u,v)\Big(\EE_{1,3}(u)-\EE_{1,2}(u)\EE_{2,3}(u)\Big)-
\gori(u,v)\EE_{1,3}(v)\,,
\end{split}
\end{equation*}
\begin{equation*}\label{CC13}
[\EE_{2,3}(v),\FF_{3,2}(u)]=\gole(u,v)\left(k_{3}(u)k_{2}(u)^{-1}-k_{3}(v)k_{2}(v)^{-1}\right)\,,
\end{equation*} 
where diagonal Gauss coordinate $k_3(u)$ due to \r{Kinv} and \r{Kinvall} is equal to
\begin{equation*}
k_3(u)=k_2(q^4u)^{-1} k^\pm_1(q^4u)k^\pm_1(q^6u)^{-1}\,.
\end{equation*}
These commutation relations allows to restore the full set of the commutation relations 
in terms of the currents for the algebra $U_q(C^{(1)}_n)$ given in section~\ref{nrCn}.

To obtain the commutation relation \r{CC11} 
one has to use  \r{TM-1} 
for  the values of the indices 
$\{i,j,k,l\}\to\{2,3,1,2\}$ and $\{i,j,k,l\}\to\{2,4,1,1\}$ which results to 
\begin{equation*}\label{C8}
\begin{split}
&f(u,v)\FF_{3,2}(u)\FF_{2,1}(v)=\frac{f(u,v)}{\fgo(u,v)}\FF_{2,1}(v)\FF_{3,2}(u)+
\frac{f(u,v)\gole(u,v)}{\fgo(u,v)}\FF_{3,1}(v)+\\
&\quad+\gle(v,u)\Big(\FF_{3,1}(u)-\FF_{3,2}(u)\FF_{2,1}(u)\Big)+
\frac{q^{-2}\gle(v,u)}{\fgo(u,v)}\FF_{4,2}(u)\,.
\end{split}
\end{equation*}
Considering the latter relation at $v=q^2 u$ we obtain 
\begin{equation}\label{C9}
\FF_{4,2}(u)=q^2\Big(\FF_{3,1}(u)-\FF_{3,2}(u)\FF_{2,1}(u)\Big)
\end{equation}
and \r{CC11}. Now one 
 can calculate the projection $\Pfp(F_3(u) F_2(u))$, where 
 dependent current $F_3(u)=-F_1(q^4u)$ is defined by \r{depcur} for $N=4$ and $\xi=q^{-6}$.
Restoring in  \r{CC11} superscripts of the matrix entries and setting  $v=q^4u$ we obtain \begin{equation*}
\FF^+_{2,1}(q^4u)\FF^\pm_{3,2}(u)=q^2\Big(\FF^\pm_{3,2}(u)\FF^\pm_{2,1}(u)-
\FF^\pm_{3,1}(u)\Big)+q^{-2}\FF^+_{3,1}(q^4u)\,.
\end{equation*}
Calculating projection $\Pfp(F_3(u) F_2(u))$ onto $U^+_f$ according to the ordering \r{order1}
one gets
\begin{equation*}
\begin{split}
\Pfp\sk{F_3(u) F_2(u)}&=-\Pfp\sk{F_1(q^4u) F_2(u)}=-\Pfp\sk{\FF^+_{2,1}(q^4u) F_2(u)}=\\
&=q^2\Big(\FF^+_{3,1}(u)-\FF^+_{3,2}(u)\FF^+_{2,1}(u)\Big)=\FF^+_{4,2}(u)\,.
\end{split}
\end{equation*}
Analogously, one can prove that 
\begin{equation*}
\Pfp\sk{F_2(u) F_1(u)}=\FF^+_{3,1}(u)\quad\mbox{and}\quad 
\Pfp\sk{F_3(u) F_2(u) F_1(u)}=\FF^+_{4,1}(u)\,.
\end{equation*}
These relations together with analogous relations for the currents $E_i(u)$ 
are base of the induction for the proof of the proposition~\ref{cvsg1} in case 
of the algebra $U_q(C^{(1)}_n)$.

\subsection{Algebra $U_q(D^{(1)}_2)$}
\label{o2n}

As above we start to consider algebra $U_q(D^{(1)}_n)$ for small $n$. 
The case $n=1$ is not representative and we begin with the case $n=2$ 
to prove that $\FF^\pm_{3,2}(u)=\EE^\pm_{2,3}(u)=0$.

Excluding term $L_{21}(v)L_{24}(u)$ from the commutation relation \r{TM-1} with set of indices 
$\{i,j,k,l\}\to\{2,3,2,2\}$ and $\{i,j,k,l\}\to\{2,4,2,1\}$ we have
\begin{equation}\label{D6}
\begin{split}
&f(u,v)\LL_{2,3}(u)\LL_{2,2}(v)-f(v,u)\LL_{2,2}(v)\LL_{2,3}(u) =\\ 
&\quad =f(v,u)\gri(v,u)\LL_{2,4}(v)\LL_{2,1}(u) - f(u,v)\gri(u,v)\LL_{2,4}(u)\LL_{2,1}(v) 
\end{split}
\end{equation}
This relation after setting $v=q^{-2}u$ and projecting onto subalgebras 
$U^\pm_f\cup U^\pm_k$  in the algebra 
$U_q(D^{(1)}_2)$ yields the equality 
\begin{equation*}\label{D7}
\FF^\pm_{3,2}(u)k^\pm_2(u)k^\pm_2(q^{-2}u)=0\,.
\end{equation*} 
Since Gauss coordinates $k^\pm_2(u)$ are invertible it results that 
\begin{equation}\label{D8}
\FF^\pm_{3,2}(u)=0.
\end{equation}
Analogously one can prove  
\begin{equation}\label{D9}
\EE^\pm_{2,3}(u)=0.
\end{equation}

In order to find relations between Gauss coordinates $\FF_{j,1}(u)$, $j=2,3,4$ 
one can consider 
the commutation relation \r{TM-1} for the values of the indices 
$\{i,j,k,l\}\to\{1,3,1,2\}$ and $\{i,j,k,l\}\to\{1,4,1,1\}$. 
Excluding term $L_{11}(v)L_{14}(u)$ we have
\begin{equation}\label{D10} 
\begin{split}
&f(u,v)\LL_{1,3}(u)\LL_{1,2}(v)=f(v,u)\LL_{1,2}(v)\LL_{1,3}(u)+\\ 
&\quad +f(v,u)\gri(v,u)\LL_{1,4}(v)\LL_{1,1}(u) - f(u,v)\gri(u,v)\LL_{1,4}(u)\LL_{1,1}(v)\,.
\end{split}
\end{equation}  
Using explicit expressions for the matrix entries $\LL_{1,j}(u)$ through Gauss 
coordinates \r{Gauss1}, multiplying both equalities by the product of $k_1(u)^{-1}k_1(v)^{-1}$ 
and using the commutation relations
\begin{equation}\label{D12}
k_1(u)\FF_{j,1}(v)k_1(u)^{-1}=f(v,u)\FF_{j,1}(v)-\gle(v,u)\FF_{j,1}(u),\quad j=2,3
\end{equation}
one can get  from \r{D10}  
\begin{equation*}\label{D13}
\begin{split}
&f(u,v)f(v,u)\Big(\FF_{2,1}(v)\FF_{3,1}(u)-\FF_{3,1}(u)\FF_{2,1}(v)\Big)=\\
&=\gle(u,v)f(v,u)\Big(\FF_{4,1}(v)+\FF_{2,1}(v)\FF_{3,1}(v)\Big)
+ \gle(v,u)f(u,v)\Big(\FF_{4,1}(u)+\FF_{3,1}(u)\FF_{2,1}(u)\Big).
\end{split}
\end{equation*}
Taking in this equality $u=q^2v$ and $u=q^{-2}v$ one can find the relations
\begin{equation}\label{D14}
\begin{split}
\FF_{4,1}(u)&=-\FF_{2,1}(u)\FF_{3,1}(u)=-\FF_{3,1}(u)\FF_{2,1}(u)\,,\\
&\FF_{2,1}(v)\FF_{3,1}(u)=\FF_{3,1}(u)\FF_{2,1}(v)\,.
\end{split}
\end{equation}
In the same way one can prove that 
\begin{equation}\label{D15}
\begin{split}
\EE_{1,4}(u)&=-\EE_{1,2}(u)\EE_{1,3}(u)=-\EE_{1,3}(u)\EE_{1,2}(u)\,,\\
&\EE_{1,2}(v)\EE_{1,3}(u)=\EE_{1,3}(u)\EE_{1,2}(v)\,.
\end{split}
\end{equation}

Equalities  \r{Finv}, \r{Einv},  \r{Finvall} and \r{Einvall} yields in this case 
\begin{equation}\label{D17}
\FF_{4,5-j}(u)=-\FF_{j,1}(u),\quad \EE_{5-j,4}(u)=-\EE_{1,j}(u),\quad j=2,3. 
\end{equation}
Using \r{TM-1} for $\{i,j,k,l\}\to\{2,2,1,3\}$ we can calculate 
\begin{equation*}\label{D18}
k_2(u)\FF_{3,1}(v)k_2(u)^{-1}=f(v\xi,u)\FF_{3,1}(v)-
q\ \gri(u,v\xi)k_1(v)\FF_{3,1}(u)k_1(v)^{-1},
\end{equation*}
where we have used \r{D8} and \r{D17}. Using now \r{D12} and  identities
\begin{equation*}
f(v\xi,u)-q\ \gri(u,v\xi) \gle(u,v)=f(v,u),\quad q\ \gri(u,v\xi) f(u,v)=-\gle(v,u)
\end{equation*}
one can find  that
\begin{equation*}\label{D19}
k_2(u)\FF_{3,1}(v)k_2(v)^{-1}=f(v,u)\FF_{3,1}(v)-\gle(v,u)\FF_{3,1}(u).
\end{equation*}
Analogously one can obtain 
\begin{equation*}\label{D20}
k_2(u)^{-1}\EE_{1,3}(v)k_2(v)=f(v,u)\EE_{1,3}(v)-\gri(v,u)\EE_{1,3}(u).
\end{equation*}

Using \r{D12} and analogous commutation relations for $\EE_{1,j}(u)$ one can calculate 
from the commutation relation \r{TM-1} at $\{i,j,k,l\}\to\{1,3,3,1\}$ that
\begin{equation*}\label{D21}
[\EE_{1,3}(v),\FF_{3,1}(u)]=\gle(u,v)\Big(k_3(u)k_1(u)^{-1}-k_3(v)k_1(v)^{-1}\Big).
\end{equation*}
The embedding theorem~\ref{thm-emb} and   commutation relations between 
Gauss coordinates  obtained for the algebra $U_q(D^{(1)}_2)$
are 
sufficient to get full set of the commutation relations for the 
algebra $U_q(D^{(1)}_n)$ in terms of the currents presented in the section~\ref{nrDn}.

\subsection{Algebra $U_q(A^{(2)}_{2})$}
\label{A22n}

$\RR\LL\LL$ realization of this algebra is given by the $\RR$-matrix \r{R-mat} with 
$\xi=-q^{-1-2n}$ and $N=2n+1$. In the same way as we investigated the algebra 
$U_q(B^{(1)}_1)$
we study  first  the algebra $U_q(A^{(2)}_{2n})$ in the simplest case $n=1$. 

Introduce the functions 
\begin{equation*}\label{fgoA2n}
\fgs(u,v)=\frac{q^{1/2}u+q^{-1/2}v}{u+v},\quad
 \ggs(u,v)=\frac{(q^{1/2}+q^{-1/2})u}{u+v}\,.
\end{equation*}
The commutation relations for the Gauss coordinates in the algebra
$U_q(A^{(2)}_{2})$  between $\FF_{2,1}(u)$, $k_1(u)$  and $\EE_{1,2}(u)$
are the same as for $U_q(B^{(1)}_1)$  (see \r{kFEB1}).   The rest relations are
\begin{equation*}\label{A22-31}
\begin{split}
    k_2(u) &\FF_{2,1}(v)k_2(u)^{-1} = f(u,v)  \frac{\fgs(v,u)}{\fgs(u,v)}  \FF_{2,1}(v) \\ 
    &+   \gle (v,u) \FF_{2,1}(u)  + (1-q)\frac{\ggs(v,u)}{\fgs(u,v)}  \FF_{2,1}(-qu) \,,
    \end{split}
\end{equation*}
\begin{equation*}\label{A22-41}
\begin{split}
    k_2(u)^{-1}&\EE_{1,2}(v) k_{2}(u) = f(u,v) \frac{\fgs(v,u)}{\fgs(u,v)} \EE_{1,2}(v)  \\ 
    &+ \gri (v,u)  \EE_{1,2}(u) + (q-1)\frac{\ggs(u,v)}{\fgs(u,v)}  \EE_{1,2}(-qu) 
\end{split}
\end{equation*}
and 
\begin{equation}\label{A22-81}
\begin{split}
 \fgs(u,v)f(v,u)& \FF_{2,1}(u) \FF_{2,1}(v) =   \fgs(v,u)f(u,v) \FF_{2,1}(v) \FF_{2,1}(u) +\\
&+ \ggs(u,v) \gle (v,u) \Big(  \FF_{2,1}(u)^2+  \FF_{2,1}(v)^2\Big)  +\\ 
 &  +\frac{1}{q+1}  [ \FF_{2,1}[0],\ggs(u,v)\FF_{2,1}(v) -\ggs(v,u) \FF_{2,1}(u)]_q\,,
\end{split}
\end{equation}
\begin{equation}\label{A22-91}
\begin{split}
 \fgs(v,u)f(u,v) &\EE_{1,2}(u) \EE_{1,2}(v) =   \fgs(u,v)  f(v,u) \EE_{1,2}(v) \EE_{1,2}(u) +\\
& + \ggs(v,u)  \gle (u,v) \Big( \EE_{1,2}(u)^2 +  \EE_{1,2}(v)^2\Big)+\\ 
&+\frac{1}{q^{-1}+1}  [ \EE_{1,2}[0],\ggs(v,u)\EE_{1,2}(v) -\ggs(u,v) \EE_{1,2}(u)]_{q^{-1}}\,.
\end{split}
\end{equation}

To prove \r{A22-81} one can use the commutation relation
\begin{equation}\label{A22-8}
\begin{split}
 &f(v,u) \FF^+_{2,1}(u) \FF^\pm_{2,1}(v) =   f(v, - q u)f(u,v) \FF^\pm_{2,1}(v) \FF^+_{2,1}(u) +\\
&\qquad +  \gle (v,u)   \FF^+_{2,1}(u)^2  +   f(v, - q u) \gri (v,u)  \FF^\pm_{2,1}(v)^2 +\\
&\qquad   +q^{-1/2} \gle(v,-q u)  \FF^+_{3,1}(u) + q^{-3/2}\gri(v,-q u) \FF^\pm_{3,1}(v)\,.
\end{split}
\end{equation}
Putting $v\to\infty$ in \r{A22-8}  one  obtains
\begin{equation}
 \FF^+_{3,1}(u) = - \sqrt q \; \FF^+_{2,1}(u)^2 - 
 \frac{\sqrt q }{q - q^{-1}} [\FF_{2,1}[0],\FF^+_{2,1}(u)]_q .
\end{equation}
Setting $u=-qv$ in \r{A22-8}  one  finds
\begin{equation*}\label{A22-12} 
\FF^+_{2,1}(-qv)\FF^\pm_{2,1}(v)=(1-q^{-1})\FF^+_{2,1}(-qv)^2-q^{-3/2}\FF^+_{3,1}(-qv)-
q^{-1/2}\FF^\pm_{3,1}(v)\,.
\end{equation*}
Using this equality one can calculate 
\begin{equation}\label{A22-13}
\Pfp(F_2(v)F_1(v))=-\sqrt{q}\Pfp(\FF^+_{2,1}(-qv)F_1(v))=\FF^+_{3,1}(v)\,,
\end{equation} 
where dependent current $F_2(v)=-\sqrt{q}F_1(-qv)$ is defined by \r{depcur2} 
for $n=1$ and $\xi=-q^{-3}$. Relation \r{A22-13} as well as analogous relation for the 
projections $\Pepm(E_1(u)E_2(u))$ are base of induction to prove proposition~\ref{cvsg1}
in case of the algebra $U_q(A^{(2)}_{2n})$.

\subsection{Algebras $U_q(A^{(2)}_{1})$ and $U_q(A^{(2)}_{3})$}
\label{A2n1}

Algebra $U_q(A^{(2)}_{2n-1})$ is defined 
 by the $\RR$-matrix \r{R-mat} with 
$\xi=-q^{-2n}$ and $N=2n$. 
In this appendix we investigate algebra  
$U_q(A^{(2)}_{2n-1})$ in two simplest cases $n=1$ and $n=2$. 

For $n=1$ $\RR$-matrix \r{R-mat} has the form 
\begin{equation*}\label{A21-1}
\RR(u,v)=\frac{u+v}{qu+q^{-1}v}
\sk{\begin{array}{cccc}
\displaystyle{\fgs(u,v)}&0&0&0\\
0&1&\displaystyle{\ggs(u,v)}&0\\
0&\displaystyle{\ggs(u,v)}&1&0\\
0&0&0&\displaystyle{\fgs(u,v)}
\end{array}},
\end{equation*}
where rational functions $\fgs(u,v)$ and $\ggs(u,v)$ are defined as follows
\begin{equation*}\label{A21-9}
\fgs(u,v)=\frac{q^2u^2-q^{-2}v^2}{u^2-v^2},\qquad \ggs(u,v)=\frac{(q^2-q^{-2})uv}{u^2-v^2}\,.
\end{equation*}

Up to overall factor this matrix coincides with the symmetric form 
of the $\RR$-matrix for the algebra $U_{q^2}(A^{(1)}_1)$. 
This results that the commutation relation for the Gauss coordinates 
$\FF_{2,1}(u)$, $\EE_{1,2}(u)$, $k_1(u)$ and $k_2(u)$ of the algebra 
$U_q(A^{(2)}_{1})$ can be written in  the form 
\begin{equation}\label{A21-2}
k_1(u)\FF_{2,1}(v)k_1(u)^{-1}=\fgs(v,u)\FF_{2,1}(v)-
\ggs(v,u)\FF_{2,1}(u)\,,
\end{equation}
\begin{equation}\label{A21-3}
k_1(u)^{-1}\EE_{1,2}(v)k_1(u)=\fgs(v,u)\EE_{1,2}(v)-
\ggs(v,u)\EE_{1,2}(u)\,,
\end{equation}
\begin{equation}\label{A21-4}
[\EE_{1,2}(u),\FF_{2,1}(v)]=\ggs(v,u)\sk{k_2(v)k_1(v)^{-1}-k_2(u)k_1(u)^{-1}}\,,
\end{equation}
\begin{equation}\label{A21-5}
k_2(u)\FF_{2,1}(v)k_2(u)^{-1}=\fgs(u,v)\FF_{2,1}(v)-
\ggs(u,v)\FF_{2,1}(u)\,,
\end{equation}
\begin{equation}\label{A21-6}
k_2(u)^{-1}\EE_{1,2}(v)k_2(u)=\fgs(u,v)\EE_{1,2}(v)-
\ggs(u,v)\EE_{1,2}(u)\,,
\end{equation}
and 
\begin{equation}\label{A21-7}
\fgs(v,u)\FF_{2,1}(u)\FF_{2,1}(v)-\ggs(v,u)\FF_{2,1}(u)^2=
\fgs(u,v)\FF_{2,1}(v)\FF_{2,1}(u)-\ggs(u,v)\FF_{2,1}(v)^2\,,
\end{equation}
\begin{equation}\label{A21-8}
\fgs(u,v)\EE_{1,2}(u)\EE_{1,2}(v)-\ggs(u,v)\EE_{1,2}(v)^2=
\fgs(v,u)\EE_{1,2}(v)\EE_{1,2}(u)-\ggs(v,u)\EE_{1,2}(u)^2\,.
\end{equation}

The commutation relations \r{A21-2}--\r{A21-4} imply certain analytical 
properties of the Gauss coordinates $\FF_{2,1}(u)$, $\EE_{1,2}(u)$
and $k_2(u)k^\pm_1(u)^{-1}$. Indeed, setting in \r{A21-2} and \r{A21-3}
$v=\pm q^{-2}u$ we obtain 
\begin{equation}\label{A21-10}
\begin{split}
k_1(u)^{-1}\FF_{2,1}(u)k_1(u)&=\pm\  \FF_{2,1}(\pm q^{-2}u)\,,\\
k_1(u)\EE_{1,2}(u)k_1(u)^{-1}&=\pm\  \EE_{1,2}(\pm q^{-2}u)
\end{split}
\end{equation}
which imply 
\begin{equation}\label{A21-12}
\FF_{2,1}(-u)=- \FF_{2,1}(u)\quad \mbox{and}\quad
\EE_{1,2}(-u)=- \EE_{1,2}(u)\,.
\end{equation}
These equalities signify that Gauss coordinates $\FF^\pm_{2,1}(u)$ and $\EE^\pm_{1,2}(u)$
are series with respect of odd powers of the spectral parameters and equalities \r{A21-10} 
are simplified to 
\begin{equation}\label{A21-100}
\begin{split}
k_1(u)^{-1}\FF_{2,1}(u)k_1(u)&=  \FF_{2,1}( q^{-2}u)\,,\\
k_1(u)\EE_{1,2}(u)k_1(u)^{-1}&= \EE_{1,2}( q^{-2}u)\,.
\end{split}
\end{equation}

On the other hand, replacing $u$ by $-u$ in \r{A21-4}, using \r{A21-12} and the fact 
that $\ggs(-u,v)=-\ggs(u,v)$ 
one obtains that 
\begin{equation}\label{A21-15}
k_2(u)k_1(u)^{-1}=
k_2(-u)k_1(-u)^{-1}
\end{equation}
which signifies that the ration $k^\pm_2(u)k^\pm_1(u)^{-1}$ are series
with respect to even powers of the spectral parameter. 
Moreover, equality \r{identalt}
together with \r{A21-100} yield in this case that $k_2^\pm(u)=k_1^\pm(-q^2u)^{-1}$. 
Together with \r{A21-15} it proves that in the algebra $U_q(A^{(2)}_1)$ both diagonal
Gauss coordinates $k^\pm_1(u)$ and $k^\pm_2(u)$ are series with respect to even powers of 
the spectral parameters.

In the case $n=2$ and according to the embedding theorem~\ref{thm-emb} 
the Gauss coordinates 
$\FF_{2,1}(u)$, $\EE_{1,2}(u)$, $k_1(u)$ and $k_2(u)$
satisfy the commutation relations in the algebra $\tilde U_q(A^{(1)}_1)$ while 
Gauss coordinates 
$\FF_{3,2}(u)$, $\EE_{2,3}(u)$, $k_2(u)$ and $k_3(u)$
satisfy the commutation relations \r{A21-2}--\r{A21-8}.
Commutation relations between these algebraically independent sets 
of the Gauss coordinates take the form 
\begin{equation}\label{A23-10}
\begin{split}
&\FF_{2,1}(v)\FF_{3,2}(u)=\fgs(u,v)\FF_{3,2}(u)\FF_{2,1}(v)-\ggs(u,v)\FF_{3,1}(v)+\\
&\quad+ \ggs(u,v)\sk{\FF_{3,1}(u)-\FF_{3,2}(u)\FF_{2,1}(u)}
+\frac{1}{[2]_q}\ \frac{\ggs(u,v)}{\gri(u,v)}\ [\FF_{2,1}[0],\FF_{3,2}(u)]_{q^{-2}}\,,
\end{split}
\end{equation}
\begin{equation*}\label{A23-15}
\begin{split}
&\EE_{2,3}(u)\EE_{1,2}(v)=\fgs(u,v)\EE_{1,2}(v)\EE_{2,3}(u)-\ggs(u,v)\EE_{1,3}(v)+\\
&\quad+ \ggs(u,v)\sk{\EE_{1,3}(u)-\EE_{1,2}(u)\EE_{2,3}(u)}
+\frac{1}{[2]_q}\ \frac{\ggs(u,v)}{\gle(u,v)}\ [\EE_{2,3}(u),\EE_{1,2}[0]]_{q^{2}}\,,
\end{split}
\end{equation*}
where 
\begin{equation*}
[2]_q=q+q^{-1}=\frac{q^2-q^{-2}}{q-q^{-1}}\,.
\end{equation*}

To prove equality \r{A23-10} one can use an equality
\begin{equation}\label{A23-7}
\begin{split}
&\FF^+_{2,1}(v)\FF^\pm_{3,2}(u)=\fgs(u,v)\FF^\pm_{3,2}(u)\FF^+_{2,1}(v)
-\ggs(u,v)\FF^+_{3,1}(v)+\\
&\quad+ \ggs(u,v)\frac{qu+q^{-1}v}{(q+q^{-1})u}\sk{\FF^\pm_{3,1}(u)
-\FF^\pm_{3,2}(u)\FF^\pm_{2,1}(u)}
-\frac{(1-q^{-2})v}{u+v}\FF^\pm_{4,2}(u)
\end{split}
\end{equation}
which helps to calculate $\Pfp(F_3(u)F_2(u))$, where $F_3(u)=-F_1(-q^2u)$
is a dependent current defined by \r{depcur} for $N=4$ and $\xi=-q^{-4}$. 
Putting $v\to\infty$ in \r{A23-7}  one obtains
\begin{equation}
 \FF^+_{4,2}(u) = \FF^+_{3,2}(u)\FF^+_{2,1}(u) -\FF^+_{3,1}(u) - \frac{ q }{q - q^{-1}} [\FF_{2,1}[0],\FF^+_{3,2}(u)]_{q^{-2}} .
\end{equation}
Setting in \r{A23-7} $v=-q^2u$ one gets 
\begin{equation*}\label{A23-28}
\FF^+_{4,3}(u)\FF^\pm_{3,2}(u)=\FF^+_{3,1}(-q^2u)+\FF^\pm_{4,2}(u)
\end{equation*} 
which implies
\begin{equation}\label{A23-100}
\Pfp\sk{F_3(u)F_2(u)}=\Pfp\sk{\FF^+_{4,3}(u)(\FF^+_{3,2}(u)-\FF^-_{3,2}(u))}=\FF^+_{4,2}(u)\,.
\end{equation}
Analogously, considering equality \r{A23-7} at $v=u$ one can prove that 
\begin{equation}\label{A23-101}
\Pfp\sk{F_2(u)F_1(u)}=\Pfp\sk{\FF^+_{3,2}(u)(\FF^+_{2,1}(u)-\FF^-_{2,1}(u))}=\FF^+_{3,1}(u)\,.
\end{equation}

Let us consider commutation relation \r{TM-1} for the values of the indices 
$\{i,j,k,l\}\to\{2,4,1,2\}$ and $\{i,j,k,l\}\to\{2,4,1,1\}$ 
to obtain commutation of the Gauss coordinates 
\begin{equation}\label{A23-29} 
\begin{split}
&f(u,v)\FF^+_{4,2}(u)\FF^\pm_{2,1}(v)=\frac{1}{f(v\xi,u)}\ \FF^\pm_{2,1}(v)\FF^+_{4,2}(u)+\\
&\quad+\frac{\gle(u,v\xi)}{qf(v\xi,u)}\Big(\FF^\pm_{2,1}(v)\FF^\pm_{3,1}(v)+
\FF^\pm_{2,1}(v)^2\FF^+_{3,2}(u)\Big)+\\
&\quad+\gle(u,v)\FF^\pm_{4,1}(v)-\gri(u,v)\Big(\FF^+_{4,1}(u)-\FF^+_{4,2}(u)\FF^+_{2,1}(u)\Big)\,.
\end{split}
\end{equation}
Taking difference of two equalities in \r{A23-29} and applying the projection $\Pfp$ 
to this difference one obtains 
\begin{equation*}\label{A23-30} 
\begin{split}
&f(u,v)\Pfp\sk{\FF^+_{4,2}(u)F_1(v)}=\gle(u,v)\FF^+_{4,1}(v)+\\
&\quad+\frac{1}{f(v\xi,u)}\ \FF^+_{2,1}(v)\FF^+_{4,2}(u)
+\frac{\gle(u,v\xi)}{qf(v\xi,u)}\Big(\FF^+_{2,1}(v)\FF^+_{3,1}(v)+
\FF^+_{2,1}(v)^2\FF^+_{3,2}(u)\Big)
\end{split}
\end{equation*}
which after setting $u=v$ implies 
\begin{equation*}\label{A23-31}
\Pfp\sk{\FF^+_{4,2}(u)F_1(u)}=\FF^+_{4,1}(u)\,.
\end{equation*}
Using latter equality and \r{A23-100} one gets 
\begin{equation}\label{A23-32}
\Pfp(F_3(u)F_2(u)F_1(u))=\Pfp(\FF^+_{4,2}(u)F_1(u))=\FF^+_{4,1}(u)\,.
\end{equation}
Equalities \r{A23-100}, \r{A23-101} and \r{A23-32} together with analogous 
relations for the currents $E_i(u)$ are base of the induction to prove 
proposition~\ref{cvsg1}.

\section{Proof of proposition~\ref{cvsg1}}\label{prf71}

We start with the proof of the first equality in \r{PrvsGc}.
Assume that it is valid in the algebra 
$U_q^{n-1}(\gaf)$. It means that in order to prove first equality 
in \r{PrvsGc} one has to prove it for the Gauss coordinates 
$\FF^+_{N,i}(u)$, $1\leq i\leq N-1$ and $\FF^+_{j,1}(u)$, $2\leq j\leq N$
using induction assumption that it is valid for all $\FF^+_{j,i}(u)$,
$2\leq i< j\leq N-1$.  

To do this we consider $\RR\LL\LL$ commutation relations 
\r{RLL} written in the form 
\begin{equation}\label{prf1}
\begin{split}
&f(u,v)f(v,u)\ (\Id\ot\LL^+(v))\cdot (\LL^-(u)\ot\Id)=\\
&\qquad= \RR_{12}(u,v)\cdot
(\LL^-(u)\ot\Id)\cdot (\Id\ot\LL^+(v))\cdot \RR_{21}(v,u)\,.
\end{split}
\end{equation}

Consider $(i,i+1)$ matrix element in the first space and $(i+1,N)$ matrix 
element in the second space of the equality \r{prf1}. After substitution in the resulting 
equality Gauss decomposition  \r{Gauss1} one can 
multiply it from the right by $k^+_{i+1}(v)^{-1}$ and 
$k^-_{i}(u)^{-1}$ and 
normal order products of Gauss coordinates according to the ordering \r{order1}.
Then one has to restrict resulting equality to subalgebra $U_f^+$
as it was described in the section~\ref{nor-ord}, multiply it by $(u-v)^3$ 
and set $u=v$. Final equality takes the form 
\begin{equation*}\label{prf22}
\begin{split}
&(u-v)\ \Big(\FF^+_{N,i+1}(v)k^+_{i+1}(v)\FF^-_{i+1,i}(u)k^+_{i+1}(v)^{-1}+\\
&\qquad+\FF^+_{N,i}(v)k^+_{i}(v)[\EE^+_{i,i+1}(v),\FF^-_{i+1,i}(u)]
k^+_{i+1}(v)^{-1}\Big)\Big|_{U_f^+}\Big|_{u=v}=0\,.
\end{split}
\end{equation*}
Using the commutation relations  
$k^+_{i+1}(v)$ and $\EE^+_{i,i+1}(v)$ with $\FF^-_{i+1,i}(u)$ which are 
different for different $U_q(\gaf)$ when $i=n+1$ 
and taking into account that restriction to subalgebra $U^+_f$ coincides 
with the action of the projection $\Pfp$ onto subalgebra $U_f$
one obtains 
the equality in $U^+_f$
\begin{equation}\label{prf6}
\FF^+_{N,i}(v)=\Pfp\Big(\FF^+_{N,i+1}(v)\Big(\FF^+_{i+1,i}(v)-\FF^-_{i+1,i}(v)\Big)\Big)=
\Pfp\Big(\FF^+_{N,i+1}(v)\ F_i(v)\Big)\,,
\end{equation} 
where $i<N-1$ for all $U_q(\gaf)$ except $U_q(D^{(1)}_n)$ and 
 \begin{equation}\label{prf7}
\FF^+_{N,i}(v)=\Pfp\Big(\FF^+_{N,i+2}(v)\Big(\FF^+_{i+2,i}(v)-\FF^-_{i+2,i}(v)\Big)\Big)
\end{equation} 
for the algebra $U_q(D^{(1)}_n)$ at $i=n-1,n$. To obtain \r{prf7} 
one can use 
 $(i,i+2)$ matrix element in the first space and $(i+2,N)$ matrix 
element in the second space of the equality \r{prf1} since $\FF^-_{n+1,n}(u)\equiv 0$ 
for the algebra $U_q(D^{(1)}_n)$.

The statement of the proposition is obviously valid for $i=N-1$ since 
$\FF^+_{N,N-1}(u)=\Pfp(F_{N-1}(u))$. Assume that it is valid for 
the Gauss coordinate $\FF^+_{N,i+1}(u)$ with $i<N-2$. Then equality \r{prf6} 
takes the form 
\begin{equation*}\label{prf8}
\FF^+_{N,i}(v)=\Pfp\Big(\Pfp\Big(F_{N-1}(v)\cdots F_{i+1}(v)\Big)\ F_i(v)\Big)\,.
\end{equation*}
The statement of the proposition~\ref{cvsg1} for the Gauss coordinate 
$\FF^+_{N,i}(v)$ is proved  by induction since projection $\Pfp$ 
possesses the property \cite{EKhP07} that for $i<j$
\begin{equation}\label{prf9}
\Pfp\Big(F_{j-1}(u)\cdots F_{i}(u)\Big)=F_{j-1}(u)\cdots F_i(u)+\ldots\,,
\end{equation}
where $\ldots$ stands for the terms annihilated by the projection $\Pfp$.

Taking in \r{prf1} matrix entries $(1,2)$  in the first space and $(2,j)$ in the second space,
using Gauss decompositions \r{Gauss1} for the matrix elements of $\LL$-operators,
multiplying resulting equality by $(u-v)^3$ and by $k^-_1(u)^{-1}k^+_2(v)^{-1}$ from the right, 
normal ordering of the Gauss coordinates and restricting to subalgebra $U_f^+$ one obtains 
after setting $u=v$ 
\begin{equation*}\label{prf3}
\FF^+_{j,1}(v)=\Pfp\Big(\FF^+_{j,2}(v)\ F_1(v)\Big)\,.
\end{equation*}
Using induction assumption for the Gauss coordinate $\FF^+_{j,2}(v)$ and \r{prf9} 
we finish proof of the first equality in   \r{PrvsGc}.

To prove the second equality in \r{PrvsGc} we have to repeat all arguments 
as above for the transpose-inverse $\LL$-operators $\hLL^\pm(u)$ using 
Gauss decomposition \r{GChL}. 
Taking the corresponding matrix elements in \r{prf1} one can find that 
\begin{equation*}\label{prf10}
\Pfm\Big(F_{j-1}(v)\ \tFF^-_{j-1,1}(v)\Big)=\tFF^-_{j,1}(v)
\end{equation*}
and 
\begin{equation*}\label{prf11}
\Pfm\Big(F_{N-1}(v)\ \tFF^-_{N-1,i}(v)\Big)=\tFF^-_{N,i}(v)\,.
\end{equation*}
By the property of projection $\Pfm$ \cite{EKhP07} 
\begin{equation*}
\Pfm\Big(F_{j}(u)\cdots F_{i}(u)\Big)=F_{j}(u)\cdots F_{i}(u)+\ldots\,,
\end{equation*}
where $\ldots$ stands for the terms which are annihilated by projection $\Pfm$
and induction assumption one can prove the second equality in the first line of 
\r{PrvsGc}. The second line  in \r{PrvsGc} can be proved similarly.\qed

\end{document}